\documentclass[conference]{IEEEtran}
\IEEEoverridecommandlockouts
\usepackage{cite}
\usepackage{amsmath,amssymb,amsfonts}
\usepackage{enumerate}
\usepackage{algorithmic}
\usepackage{pifont}
\usepackage{bbding}
\usepackage{utfsym}
\usepackage{fontawesome}
\usepackage{bbm}
\usepackage{bbding}
\usepackage{url}
\usepackage{graphicx}
\usepackage{comment}
\usepackage{textcomp}
\usepackage{subfig}
\usepackage{multirow}
\usepackage{weiwAlgorithm}
\usepackage{booktabs,tabularx,threeparttable}
\usepackage{xcolor}
\def\BibTeX{{\rm B\kern-.05em{\sc i\kern-.025em b}\kern-.08em
    T\kern-.1667em\lower.7ex\hbox{E}\kern-.125emX}}

\newcommand{\myblue}[1]{{\color{blue} #1}\xspace}

\newcommand{\Cmt}[1]{
\vspace{0.6mm} \tcp*[h]{\underline{#1}}\\ \vspace{0.6mm}
}
\newcommand{\myparagraph}[1]{\vspace{0.5mm} \noindent \textbf{#1}.}

\begin{document}

\title{Efficient Dynamic Attributed Graph Generation
}

\DeclareRobustCommand*{\IEEEauthorrefmark}[1]{%
\raisebox{0pt}[0pt][0pt]{\textsuperscript{\footnotesize\ensuremath{#1}}}}

\author{\IEEEauthorblockN{Fan Li\IEEEauthorrefmark{1},
Xiaoyang Wang\IEEEauthorrefmark{1 \ast \thanks{\textsuperscript{$\ast$}Corresponding author}},
Dawei Cheng\IEEEauthorrefmark{2,3}, Cong Chen\IEEEauthorrefmark{2}, Ying Zhang\IEEEauthorrefmark{4 \ast}, and Xuemin Lin\IEEEauthorrefmark{5}}
\IEEEauthorblockA{\IEEEauthorrefmark{1}\text{University of New South Wales, Australia}}
\IEEEauthorblockA{\IEEEauthorrefmark{2}\text{Tongji University, China}
\IEEEauthorrefmark{3}\text{Shanghai AI Laboratory, China}}
\IEEEauthorblockA{\IEEEauthorrefmark{4}\text{Zhejiang Gongshang University, China} \IEEEauthorrefmark{5}\text{Shanghai Jiao Tong University, China}}
\text{\{fan.li8, xiaoyang.wang1\}@unsw.edu.au, \{dcheng, clairchen\}@tongji.edu.cn,}\\
\text{ying.zhang@zjgsu.edu.cn, xuemin.lin@sjtu.edu.cn}
}

\maketitle

\begin{abstract}
Data generation is a fundamental research problem in data management due to its diverse use cases, ranging from testing database engines to data-specific applications. However, real-world entities often involve complex interactions that cannot be effectively modeled by traditional tabular data. Therefore, graph data generation has attracted increasing attention recently.
Although various graph generators have been proposed in the literature, there are three limitations: $i)$ They cannot capture the co-evolution pattern of graph structure and node attributes. $ii)$ Few of them consider edge direction, leading to substantial information loss. $iii)$ Current state-of-the-art dynamic graph generators are based on the temporal random walk, making the simulation process time-consuming. 
To fill the research gap, we introduce VRDAG, a novel variational recurrent framework for efficient dynamic attributed graph generation. 
Specifically, we design a bidirectional message-passing mechanism to encode both directed structural knowledge and attribute information of a snapshot. 
Then, the temporal dependency in the graph sequence is captured by a recurrence state updater, generating embeddings that can preserve the evolution pattern of early graphs. 
Based on the hidden node embeddings, a conditional variational Bayesian method is developed to sample latent random variables at the neighboring timestep 
for new snapshot generation. 
The proposed generation paradigm avoids the time-consuming path sampling and merging process in existing random walk-based methods, significantly reducing the synthesis time. Finally, comprehensive experiments on real-world datasets are conducted to demonstrate the effectiveness and efficiency of the proposed model.
\end{abstract}

\section{Introduction}
\label{sec:introduction}


In industrial practice, a DBMS is required to adequately test its database engine and applications with representative data and workloads that accurately mimic the data processing environments at customer deployments~\cite{tay2011data,sanghi2018hydra,sanghi2023synthetic}. However, due to privacy restrictions, actual data and workloads from enterprises are not usually available to database vendors~\cite{lu2014generating}. A popular and effective solution is data generation, such as relational data generation~\cite{arasu2011data,rabl2015just,sanghi2018hydra,gilad2021synthesizing,yang2022sam}. The synthetic data, which can capture desired schematic properties and statistical profiles of the original data, has been widely used in the assessment of the performance impacts of engine upgrades, business data masking, and application-specific benchmarking. 

Although significant efforts have been made towards the generation of relational data, conventional tabular data cannot effectively model the complex relationships among entities in real-world applications. As a ubiquitous data structure, graphs are adopted to model these data interactions~\cite{wu2024efficient,wang2024efficient,li2024adarisk}, such as web links~\cite{shin2019sweg} and friendships among people~\cite{ching2015one}. 
In particular, there are several strong motivations for studying graph generation in data management: (1) A lack of realistic graphs for evaluating the performance of the graph processing systems~\cite{park2017trilliong}. (2) Synthetic graphs that capture the distribution of the real-life network can be used in many social analysis tasks~\cite{wang2021fastsgg,xiang2022efficient}, such as community detection~\cite{newman2004finding} and network representations~\cite{perozzi2014deepwalk}. (3) The simulated graph anonymizes node entities and their link relationships, preventing information leakage of private data~\cite{xiang2021general}.

Nevertheless, most existing studies primarily focus on the generation of static graph structures, overlooking the dynamic nature of networks and rich node attribute information, which are two important characteristics present in most real-world graphs~\cite{li2017attributed}. 
For instance, we consider a graph-based application: fraud detection on financial transaction graphs. Potential fraudsters are subject to the limited time of activities and would change their attacking strategies over time, leading to the co-evolution of topology connection (e.g., change of transaction objects) and node attributes (e.g., change of location, transaction amount, etc)~\cite{cheng2020graph}. As the network data with user profiles and transaction records is sensitive, we cannot directly access the source data from financial institutions. Therefore, the high-quality synthetic data, which preserves the properties of the real network, can greatly facilitate the graph mining community, enabling the analysis of dynamic node behaviors and subsequently the design of effective prediction models. This, in turn, will help prevent financial losses caused by high-risk transactions.
Thus, ignoring the dynamic feature and node information of graphs will cause significant information loss, and 
it is necessary to study how to generate realistic dynamic attributed graphs effectively and efficiently.

In the literature, extensive studies have been conducted on static graph generation. Traditional approaches, such as random graph models~\cite{watts1998collective,albert2002statistical} and block-based models~\cite{karrer2011stochastic,kolda2014scalable} generate realistic graphs with predefined rules. To further generate network data for benchmarking the graph processing system and analyzing social networks, a line of efficient and scalable generators~\cite{hadian2016roll,park2017trilliong,wang2021fastsgg} has been introduced.
However, these methods rely on specific statistical assumptions in simulation and fail to capture structure distribution directly from observed graphs. To address this, a variety of data-driven deep generators including the VAE-based~\cite{kipf2016variational,grover2019graphite}, RNN-based~\cite{you2018graphrnn,liao2019efficient}, and GAN-based methods~\cite{bojchevski2018netgan,xiang2022efficient} have been proposed. For instance, CPGAN~\cite{xiang2022efficient} uses a unified GAN framework to efficiently simulate networks that can preserve the community structure in real-world graphs.
However, few of the above methods can synthesize node attributes. ANC~\cite{largeron2015generating} models the attribute distribution with a simple normal distribution, while the latest state-of-the-art, GenCAT~\cite{maekawa2023gencat}, can deal with user-specified attribute distribution.

Recently, there are some works trying to learn the rules that govern the evolution of networks over time and generate dynamic graphs~\cite{perra2012activity,michail2016introduction,zeno2021dymond,zhou2020data,gupta2022tigger}. Early studies focus on modeling the evolution of edges or nodes over time~\cite{perra2012activity,holme2013epidemiologically}, ignoring the high-order structure within graphs. To overcome this, Dymond~\cite{zeno2021dymond} considers the dynamic changes in graph structure using temporal motif activity. As the first data-driven dynamic graph generator, TagGen~\cite{zhou2020data} proposes a random walk sampling strategy to
jointly extract the structural and temporal context information directly from the observed graphs. To better model the generative distributions of temporal graphs
and retain continuous-time information, TGGAN~\cite{zhang2021tg} learns the representations of temporal graphs via truncated temporal walks.TIGGER~\cite{gupta2022tigger} further extends the temporal random walk by combining temporal point processes with autoregressive modeling, achieving state-of-the-art performance.

However, the above methods still suffer from several limitations in our task. 
$i)$ Existing generators either focus on static graph generation or overlook attribute synthesis in dynamic graph generation. Specifically, current static attributed graph generators~\cite{largeron2015generating,maekawa2023gencat} cannot capture evolution patterns in graph sequence. Besides, they can only support attribute generation based on given simple prior distributions, and fail to learn unknown distributions directly from observed data. For dynamic graph generators, traditional methods like Dymond primarily focus on simulating the behavior of edges or substructures, ignoring attribute information. The state-of-the-art deep generative models~\cite{zhou2020data,zhang2021tg,gupta2022tigger} cannot preserve the attribute information during temporal random walk sampling.
$ii)$ Most graph generation models only consider undirected networks, despite the fact that directed edges, which represent the flow of information, contain valuable relational knowledge in many real-world graphs. For instance, in a guaranteed-loan network, the directed link represents the guarantee relation from the source guarantor node to the destination borrower entity~\cite{cheng2021efficient}. It is crucial to capture the information flow pattern when synthesizing complex social networks. 
$iii)$ Existing data-driven dynamic generative models are extremely time-consuming. TagGen, TGGAN, and TIGGER are all random walk-based methods, and the generation quality relies on sampling a large number of candidate temporal random walks. Moreover, further candidate path discrimination and path merging also lead to heavy computation burdens. It is desirable to develop a new learning-based method, which can achieve a better trade-off between
the efficiency and the simulation quality.

To address these issues, we introduce VRDAG (\textbf{\underline{V}}ariational \textbf{\underline{R}}ecurrent \textbf{\underline{D}}ynamic \textbf{\underline{A}}ttributed Graph \textbf{\underline{G}}enerator). To the best of our knowledge, we are the first to investigate the problem of generating dynamic topology and attribute data simultaneously via a data-driven manner. Specifically, we utilize a bi-flow graph encoder to preserve the directional message flow knowledge and attribute information of the current snapshot into node embeddings. Then, a GRU-based submodule is utilized to capture time dependency in the generated sequence and output updated hidden node states. 
Next, we parameterize a prior distribution that can sample flexible latent variables at the neighboring timestep based on the hidden states of early graphs. These high-level latent random variables can effectively capture dependencies between and within the topological
and node attribute evolution processes. The prior distribution is learned by minimizing its discrepancy with the posterior distribution conditioned on ground-truth data. After sampling the temporal latent variables, we concatenate them with historical hidden node embeddings and send them to a factorized generator for decoding. The generator consists of a MixBernoulli sampler and an attribute decoder. The former models the probability of generating directed edges via a mixture of Bernoulli distributions, while the latter subsequently synthesizes node attributes conditioned on the newly generated structure and historical states. We repeat the above snapshot generation process and finally obtain the whole graph sequence. 
Different from the generation frameworks based on the temporal random walk, we encode the temporal state of the sequential graph into latent representations at each timestep, without the need to sample tons of candidate temporal paths. Moreover, our attributed graph decoder can generate the new snapshot in a one-shot way, whereas previous deep generators require merging the sampled paths and continuously modifying the generated graph. 
These advantages significantly accelerate the generation speed of our model. Besides, unlike the current static graph-based VAE, which fails to capture the evolution patterns of attribute data, our variational architecture with a recurrent updater generates evolving attributes that can be further used for synthesizing new links. This capability enables VRDAG to model the co-evolution patterns observed in real-world dynamic attributed graphs. Extensive experimental results demonstrate the superior effectiveness and efficiency of VRDAG in dynamic attributed graph generation compared with state-of-the-art baselines. 
The main contributions of the paper are summarized as follows.

\begin{itemize}
  \item We propose VRDAG, an efficient variational recurrent framework for dynamic attributed graph generation. 
  To the best of our knowledge, we are the first to investigate generating dynamic graphs with attribute data using a data-driven approach. Our model can directly capture the co-evolution patterns of structural information and node attributes from the realistic graph. 
  \item To preserve directed structural knowledge and node attribute information, a bi-flow message passing mechanism is proposed. In addition, a recurrent variational auto-encoder is introduced to model temporal dependencies between the latent node embeddings across timesteps. Moreover, we factorize the decoding process into topology generation based on the Mixture Bernoulli and attribute synthesis conditioned on the generated structure.
  \item Our approach is extensively evaluated on five public benchmark datasets and a real-life guaranteed-loan network. Experimental results show that VRDAG achieves state-of-the-art effectiveness and efficiency in dynamic attributed graph generation. In particular, our recurrent framework improves the generation efficiency by up to 4 orders of magnitude.
  
  
\end{itemize}

\textit{Note that, due to the limited space, some experimental settings, additional evaluations, and related works can be found in our online appendix~\cite{onlineapp}.}

\section{Preliminary}
\label{sec:pre}

\subsection{Problem Definition}

A dynamic attributed graph $G = \{G_{t}(V_{t},E_{t},X_{t})| t=1,2,...,T\}$ is a sequence of attributed graph snapshots $G_{t}(V_{t},E_{t},X_{t})$ across timesteps $t=1,2,...,T$. $V$ denotes all the unique nodes in the dynamic graph $G$ (i.e., $V=\cup_{t=1}^{T}V_{t}$) and $|V|=N$. Then, the snapshot in timestep $t$ can also be reformulated as $G_{t}(V, E_{t}, X_{t})$. In this way, the structure evolution can be reflected as the change of edges $E_{t}$ over time. 
The attribute evolution can be reflected as the change of attribute matrix $X_{t} \in R^{N \times F}$ across time, where $F$ denotes the dimension of the attribute vector $x_{v}$ for each node $v$. As graph structure $(V,E_{t})$ can be modeled as adjacency matrix $A_{t} \in R^{N \times N}$, the dynamic attributed graph can be reformulated as $G = \{G_{t}(A_{t},X_{t})| t=1,2,...,T\}$.

\noindent \textbf{Problem statement}. Given a dynamic attributed graph $G = \{G_{t}(A_{t}, X_{t})| t=1,2,..., T\}$ sampled from a complex joint distribution of graph topology, node attributes and temporal properties, we aim to estimate the underlying distribution by training a graph generator $p_{\theta}(G)$ to maximize the likelihood of generating $G$. Subsequently, we can simulate new dynamic attributed graph $\Tilde{G}=\{\Tilde{G}_{t}(\Tilde{A}_{t}, \Tilde{X}_{t})| t=1,2,..., T\}$.

\subsection{Variational AutoEncoder}

Variational autoencoder (VAE)~\cite{kingma2013auto} has emerged as a powerful probabilistic generative model for approximating the complex distribution of data space $p(x)$. It introduces a vector of latent variable $z$, which can be sampled from a prior distribution $p(z)$. Using the sampled $z$, we can explicitly reconstruct $x$ by employing a conditional distribution $p_{\theta}(x|z)$ that depends on $z$, as follows:
\begin{equation}
    p(x) = \int p_{\theta}(x|z)p(z)dz
\end{equation}
where $p_{\theta}(x|z)$ is modeled as a nonlinear neural network. The objective of the model is to maximize the likelihood of data $p(x)$. 
However, the introduction of a highly nonlinear mapping from $z$ to $x$ leads to the intractable inference of the true posterior $p(z|x)$. Consequently, the VAE proposes a variational approximation $q_{\psi}(x|z)$ as a replacement for the posterior and maximizes a variational lower bound, given by:
\begin{equation}
    \mathrm{log}p(x) \geq -\mathrm{KL}(q_{\psi}(x|z)||p(z))+\mathrm{E}_{q_{\psi}(x|z)}[\mathrm{log}p_{\theta}(x|z)]
\end{equation}
where $\mathrm{KL}(\cdot,\cdot)$ denotes the Kullback-Leibler (KL) divergence~\cite{hershey2007approximating}. In vanilla VAE, $p(z)$ is commonly chosen as a standard Gaussian distribution, while $q_{\psi}(z|x)$ and $p_{\theta}(x|z)$ are parameterized as neural networks. These two networks are jointly trained using the reparameterization trick \cite{kingma2013auto}.

\section{The Proposed Approaches}
\label{sec:method}

\begin{figure*}
\centering
\includegraphics[width=0.8\linewidth]{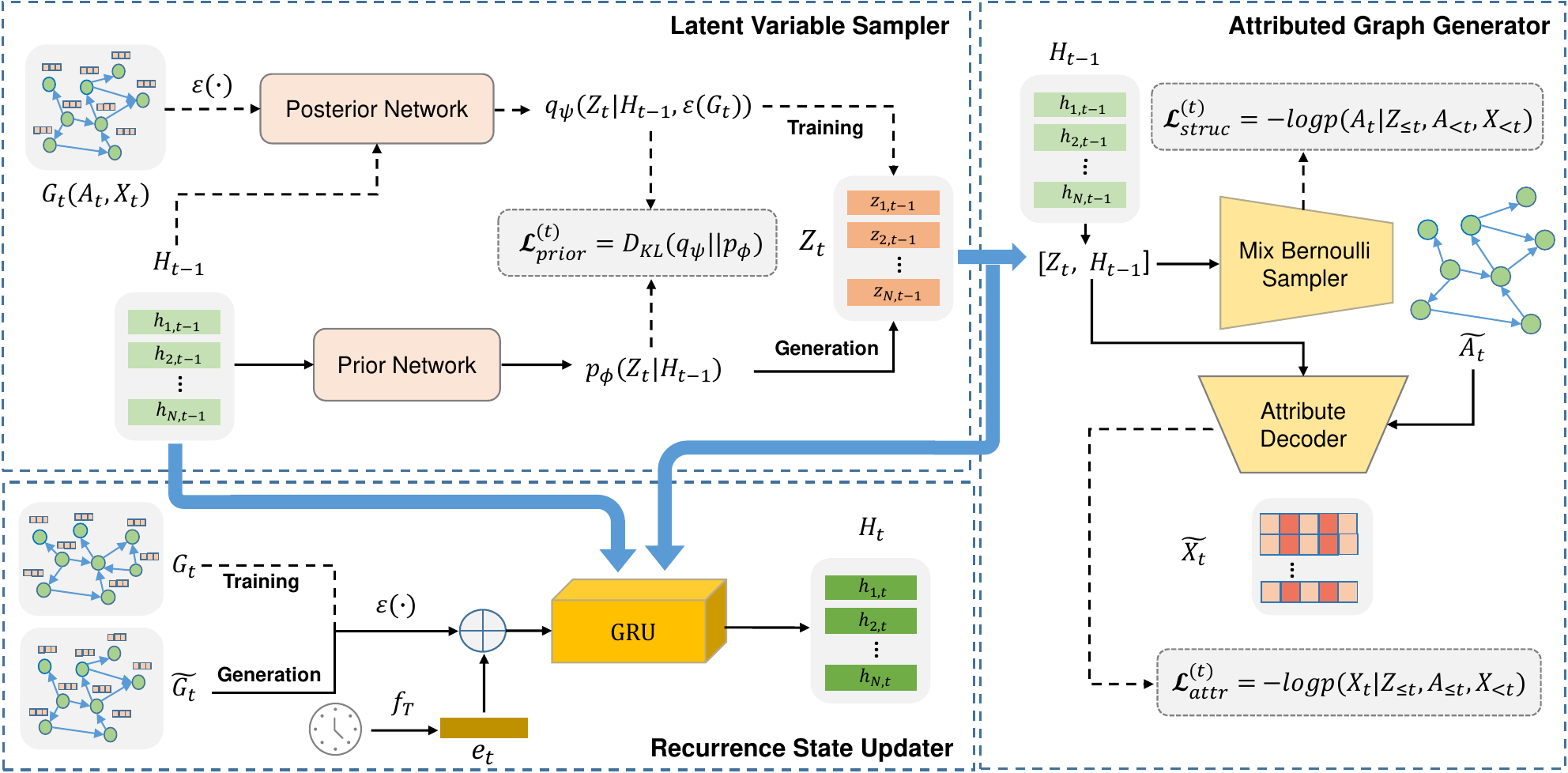}
\caption{Overview of the VRDAG framework}\label{fig:arch}
\end{figure*}

In this section, we first give the overall pipeline of our VRDAG. Then, we present the technical details of the three sub-modules in our generation pipeline, including latent variable sampler, attributed graph generator, and recurrence state updater. Next, we discuss the joint optimization strategy in our model and present the overall generative process. Finally, we give a detailed complexity analysis.

\subsection{Overall Pipeline} 

Consider a dynamic attributed graph $G = \{G_{t}(A_{t}, X_{t})| t=1,2,..., T\}$, our goal is to learn a deep graph generator $p_{\theta}$ that captures temporal dependencies between snapshots in the original graph. To achieve this, we propose to model each snapshot $G_{t}$ with a graph-based VAE conditioned on the historical hidden node states $H_{t-1} \in \mathbb{R}^{N \times d_{h}}$ updated by a GRU-based recurrent model, where $d_{h}$ is the dimensionality of $H_{t-1}$. A bi-flow GNN is developed to encode both structural and attribute information into $H_{t-1}$. The above design allows our architecture to sample temporal latent variables $Z_{t} \in \mathbb{R}^{N \times d_{z}}$ for nodes at each timestep $t$, where $d_{z}$ denotes the dimensionality of $Z_{t}$. In our implementation, we treat the sampling distribution as a learnable prior distribution $p_{\phi}$ conditioned on $H_{t-1}$. The sampled high-level node latent variables capture the co-evolution patterns of topology and attribute data and will be sent into the decoder for snapshot generation at each timestep. The decoder is an attributed graph generator that consists of a MixBernoulli sampler and an attention-based attribute simulator for generating structure and attribute, respectively.

During training, with the original graph $G$ as input, we aim to learn the parameterized prior distribution $p_{\phi}$ by minimizing its discrepancy with the posterior distribution $q_{\psi}$ that can be modeled by maximizing the likelihood of generating the original graph sequence $G$. Specifically, we optimize structure and attribute reconstruction tasks to approximate posterior distribution and train our attributed graph generator for decoding $Z_{t}$.
During inference, with the parameterized prior distribution and attributed graph decoder trained on the original graph $G$, we sample the temporal latent variables from $p_{\phi}$ and decode the new snapshot $\Tilde{G}_{t}$ recurrently. Finally, we obtain the generated graph sequence $\Tilde{G}=\{\Tilde{G}_{t}(\Tilde{A}_{t}, \Tilde{X}_{t})| t=1,2,..., T\}$. The overall architecture of VRDAG is illustrated in Figure~\ref{fig:arch}.

\subsection{Latent Variable Sampler}

To learn the prior distribution used to sample temporal latent variables $Z_{t}$, we propose to approximate it by performing maximum a posteriori (MAP) inference on the parameters during training. Therefore, we then design a posterior distribution that takes into account not only the previous state $H_{t-1}$ but also the ground truth snapshot $G_{t}(A_{t}, X_{t})$ at timestep $t$, we provide more details on modeling the above two distributions.

\subsubsection{Prior Distribution}

The prior latent distribution is applied in the generation stage where the model simulates the new dynamic graph autoregressively from scratch. Thus, the prior distribution is only conditioned on the hidden state $H_{t-1}$ updated by recurrence updater as $p_{\phi}(Z_{t}|H_{t-1})$. We can find that $H_{t-1}$ is the function of $G_{< t}$ and $Z_{< t}$. Therefore, the conditional prior distribution can be reformulated as $p_{\phi}(Z_{t}|G_{<t}, Z_{<t})$. This suggests that our variational recurrent model can capture the time dependencies among latent random variables.
Inspired by VGAE~\cite{kipf2016variational}, we decompose the prior latent distribution as:

\begin{equation}
    \begin{split}
    p_{\phi}(Z_{t}|G_{<t}, Z_{<t}) & = p_{\phi}(Z_{t}|H_{t-1}) \\
    p_{\phi}(Z_{t}|H_{t-1}) & = \prod_{i=1}^{n} p_{\phi}(z_{i,t}|h_{i,t-1}) \\
    p_{\phi}(z_{i,t}|h_{i,t-1}) & = \mathcal{N}(z_{i,t}|\mu_{i,t},diag(\sigma_{i,t}^{2}))
    \end{split}
\end{equation}
where $z_{i,t}$ is the latent embedding for node $i$ at the current step $t$, $h_{i,t-1}$ denotes the historical hidden state of node $i$. 

Specifically, we choose the multi-layer perceptron (MLP) as the prior VAE network and use the reparameterization trick to express $z_{i,t}$ as a deterministic variable. This sampling process for latent variables can be formulated as:

\begin{equation}
    \begin{split}
    \mu_{i,t} & = \mathbf{W^{\mu}_{prior}}(\omega(\mathbf{W_{prior}}h_{i,t-1})) \\ 
    \sigma_{i,t} & = \mathbf{exp}(\mathbf{W^{\sigma}_{prior}}(\omega(\mathbf{W_{prior}}h_{i,t-1}))) \\
    z_{i,t} & = \mu_{i,t}+\epsilon \cdot \sigma_{i,t},\epsilon \sim \mathcal{N}(0,1) \\
    \end{split}
\end{equation}  
where $\mathbf{W^{\mu}_{prior}}$, $\mathbf{W^{\sigma}_{prior}} \in \mathbb{R}^{d_{z} \times d_{h}}$, $\mathbf{W_{prior}} \in \mathbb{R}^{d_{h} \times d_{h}}$ are learnable weight matrices in the prior network. $\omega(\cdot)$ denotes a nonlinear transformation of the Leaky
Rectified Linear Unit. $\mathbf{exp}(\cdot)$ is exponential operation. $d_{z},d_{h}$ are the size of latent variables and hidden states, respectively. 

\subsubsection{Posterior Distribution}

During training, the model is trained by maximizing the likelihood of generating input dynamic attributed graph $G$. Thus, the posterior latent distribution $q_{\psi}$ is conditioned not only on hidden state $H_{t-1}$ but also on the observed snapshot $G_{t}(A_{t}, X_{t})$ at timestep $t$ to sample temporal latent variables for graph reconstruction task. To learn the prior distribution, we try to minimize its discrepancy with the posterior distribution.

\begin{figure}[tb!]
\centering
\includegraphics[width=0.8\linewidth]{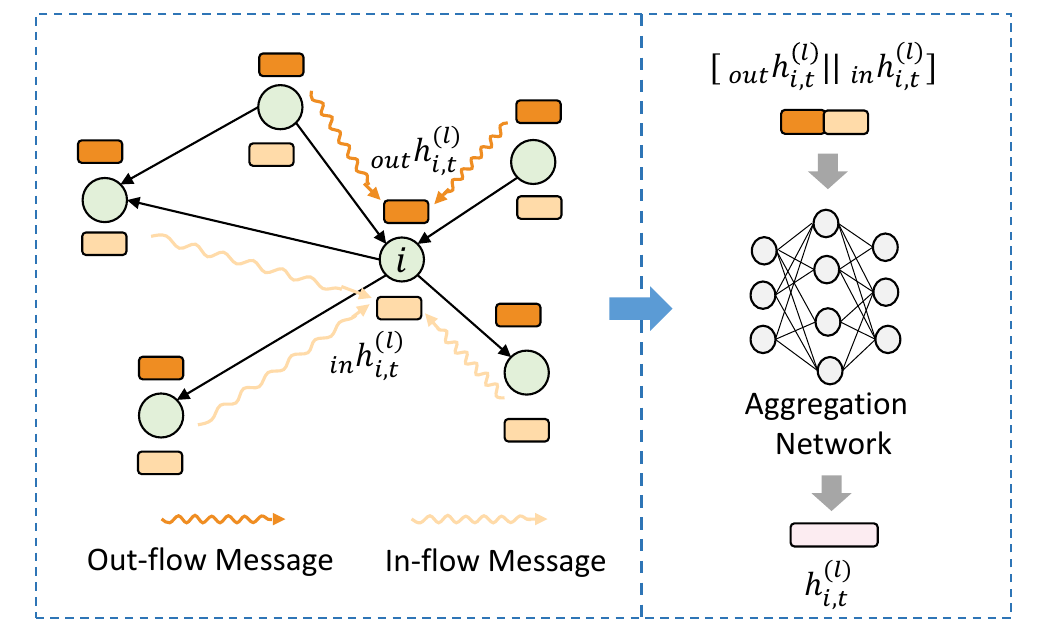}
\caption{The bi-flow message passing layer}
\label{fig:biencoder}
\end{figure}

Before modeling the latent space, we hope to incorporate knowledge of the message flow dynamics and node attributes into the node embeddings. To address this, a bi-flow graph encoder $\varepsilon$ is proposed. In particular, we divide the node state into in-flow state $_{in}h_{i,t}$ and out-flow state $_{out}h_{i,t}$, which denote the node status in different information-flow neighborhoods. To aggregate structural and attribute knowledge from two neighborhoods, we adopt GIN~\cite{xu2018powerful}, a GNN variant with strong discriminative power. The bidirectional message-passing mechanism can be formulated as:

\begin{equation}
    \begin{split}
    _{in}h_{i,t}^{(l)} & = f_{in}^{(l)}\Big((1+\epsilon_{in}^{(l)}) \cdot h_{i,t}^{(l-1)}+\sum_{v_{j} \in N_{in}(v_{i})} h_{j,t}^{(l-1)}\Big) \\ 
    _{out}h_{i,t}^{(l)} & = f_{out}^{(l)}\Big((1+\epsilon_{out}^{(l)}) \cdot h_{i,t}^{(l-1)}+\sum_{v_{j} \in N_{out}(v_{i})} h_{j,t}^{(l-1)}\Big) \\
    \end{split}
\end{equation}
where $_{in}h_{i,t}^{(l)}, _{out}h_{i,t}^{(l)}$ are in-flow and out-flow state for node $v_{i}$ at timestep $t$ in $l$-th encoder layer. $f_{in}^{(l)},f_{out}^{(l)}$ are MLPs for in-flow encoder and out-flow encoder in $l$-th layer, respectively. $N_{in}(v_{i})$ and $N_{out}(v_{i})$ represent the sets of in-neighborhood and out-neighborhood of node $v_{i}$. In each layer, we apply a node state aggregator $f_{agg}$ which takes the concatenation of in and out node embedding as input and outputs the hop-level node state $h_{i,t}^{(l)}$. The aggregation operation can be written as:
\begin{equation}
    h_{i,t}^{(l)} = f_{agg}(\left[_{in}h_{i,t}^{(l)}||_{out}h_{i,t}^{(l)}\right])
\end{equation}
where $f_{agg}$ is an $\mathrm{MLP}$ that shares weights across different layers. $||$ is the concatenation operator. The above mechanism is illustrated in Fig~\ref{fig:biencoder}. After multi-layer bi-flow message passing, we use the jump-connection technique~\cite{xu2018representation} to integrate the hop-level node states from each layer, which contain hierarchical flow information, into the final node representation $\varepsilon(v_{i,t})$ with dimensionality $d_{\varepsilon}$:
\begin{equation}
    \varepsilon(v_{i,t})=f_{pool}\big(h_{i,t}^{(1)}, h_{i,t}^{(2)},...,h_{i,t}^{(L)}\big)
\end{equation}
where $f_{pool}$ is the pooling function that is also an MLP. 

In a similar way to the prior distribution, the posterior distribution can be reformulated as $q_{\psi}(Z_{t}|G_{\leq t}, Z_{<t})$ which can be further written as:

\begin{equation}
    \begin{split}
    q_{\psi}(Z_{t}|G_{\leq t}, Z_{<t})&=q_{\psi}(Z_{t}|G_{t},H_{t-1}) \\
    q_{\psi}(Z_{t}|G_{t},H_{t-1}) & = \prod_{i=1}^{n} q_{\psi}(z_{i,t}|\varepsilon(v_{i,t}),h_{i,t-1}) \\
    q_{\psi}(z_{i,t}|\varepsilon(v_{i,t}),h_{i,t-1}) & = \mathcal{N}(z_{i,t}|\mu_{i,t},diag(\sigma_{i,t}^{2}))
    \end{split}
\end{equation}

We also leverage an MLP to build the posterior network and model the latent distribution as a Gaussian distribution:

\begin{equation}
    \begin{split}
        \mu_{i,t},\sigma_{i,t} & = \mathrm{NN_{post}}(\varepsilon(v_{i,t}),h_{i,t-1}) \\
        z_{i,t} & = \mu_{i,t}+\epsilon \cdot \sigma_{i,t},\epsilon \sim \mathcal{N}(0,1) \\
    \end{split}
\end{equation}
where $\mathrm{NN_{post}}$ denotes the posterior network that has a similar parameter architecture as the prior network in Eq. (4).
 
\subsection{Attributed Graph Generator}

Existing attributed graph generators can only synthesize attributes with pre-defined statistical distributions and ignore the inherent dependency between topology and attributes in the generation process. Besides, they fail to handle temporal information as they focus on static generation. To this end, we propose a data-driven factorized attributed graph generator for the decoding process in this subsection.

In real-world scenarios, node attributes and graph structure are co-evolving and mutually influence each other over time~\cite{wang2021modeling}. For example, consider a co-author network, forming a link (i.e., a new collaboration) extends the research scope (e.g., having new research topics) and increases the impact of the authors (e.g., having a
higher value of h-index). This attribute evolution can in turn help authors to establish new connections~\cite{wang2019tube}.
Motivated above, we design a recurrent generation method that can effectively model this co-evolution process.

For each timestep, it is natural for us to define the joint distribution across time $p_{\varphi}(A_{t}, X_{t}|Z_{\leq t}, A_{<t}, X_{<t})$ based on $p_{\varphi}(A_{t}, X_{t}|Z_{t}, H_{t-1})$ since that $H_{t-1}$ is the function of $G_{<t}(A_{<t}, X_{<t})$ and $Z_{<t}$. To decompose the joint distribution, we use a dependency-aware factorization method:
\begin{equation}
\begin{aligned}
& p_{\varphi}(A_{t}, X_{t}|Z_{\leq t}, A_{<t}, X_{<t}) \\
       & = p_{\varphi}(X_{t}|Z_{\leq t}, A_{\leq t}, X_{<t}) \cdot p_{\varphi}(A_{t}|Z_{\leq t}, A_{<t}, X_{<t})
\end{aligned}
\end{equation}
where we first generate graph topology based on the latent representation $Z_{\leq t}$ and historical snapshots $G_{< t}(A_{<t}, X_{<t})$, then we decode node attributes conditioned on the generated graph structure. 
We point out that while VRDAG follows a structure-first, attribute-later approach at each timestep, the generation of structure and attributes alternates and mutually influences each other, aligning with the principles of co-evolution. Specifically, in each synthesis step, VRDAG first generates new links based on hidden node states which capture not only historical structure information but also attributes information. Then, based on newly generated topology and historical information, we update the node attributes which will be used to establish new links at the next step.

To achieve the snapshot generation, our attributed graph generator consists of two submodules: the MixBernoulli sampler for structure generation and the attribute decoder for attribute generation. For simplicity, we denote $S_{t}=[Z_{t}||H_{t-1}]$ as the concatenation of $Z_{t}$ and $H_{t-1}$, where $s_{i,t}=[z_{i,t}||h_{i,t-1}]$.

\subsubsection{MixBernoulli Sampler}

For topology generation, we model the probability of generating the adjacency list of node $v_{i}$ at timestep $t$ via a mixture of Bernoulli distributions as:

\begin{equation}
    \begin{split}
        p_{\varphi}(\Tilde{A}_{t}|Z_{\leq t}, A_{<t}, X_{<t}) & = \prod_{i=1}^{n}p_{\varphi}(\Tilde{A}_{i,t}|Z_{\leq t}, A_{<t}, X_{<t})  \\ 
        p_{\varphi}(\Tilde{A}_{i,t}|Z_{\leq t}, A_{<t}, X_{<t}) & = \sum_{k=1}^{K}\alpha_{k,i}\prod_{1\leq j \leq N}\theta_{k,i,j} \\ 
        \alpha_{1,i},\alpha_{2,i},...,\alpha_{K,i} & = \mathrm{Softmax}(\sum_{1 \leq j \leq N} f_{\alpha}(s_{i,t}-s_{j,t})) \\
        \theta_{1,i,j},\theta_{2,i,j},...,\theta_{K,i,j} & = \mathrm{Sigmoid}(f_{\theta}(s_{i,t}-s_{j,t}))
    \end{split}
\end{equation}
where $\Tilde{A}_{i,t}$ is the generated adjacency vector of node $v_{i}$ at timestep $t$, and $\Tilde{A}_{t}$ is the whole simulated adjacency matrix. $f_{\alpha},f_{\theta}: \mathbb{R}^{d_{z}+d_{h}} \rightarrow \mathbb{R}^{K}$ are MLPs with LeakyRELU activation function. $K$ denotes the number of mixture components. When $K>1$, the generation of the individual edge is not independent, and the choice of mixture component $\alpha_{k, i}$ allows the model to sample from the best distribution based on the entire graph state, which also makes the different rows of adjacency can be computed in parallel, unlike the traditional autoregressive scheme~\cite{liao2019efficient}, which is only conditioned on the partially generated components. 

\subsubsection{Attribute Decoder}


Given the generated adjacency $\Tilde{A}_{t}$ and node states $S_{t}$, we adopt a vanilla graph attention network ~\cite{velivckovic2018graph} to simulate a message passing on the generated graph structure and then produce node attributes:

\begin{equation}
    \Tilde{X}_{t} = \sigma(\mathrm{MLP}_{attr}(\mathrm{GAT}(S_{t},\Tilde{A_{t}})))
\end{equation}
where $\Tilde{X}_{t} \in \mathbb{R}^{N \times F}$ is the matrix of decoded attributes. The attentive graph neural network is denoted as $\mathrm{GAT}(\cdot,\cdot)$ and the MLP is represented as $\mathrm{MLP}_{attr}$. $\sigma(\cdot)$ denotes a nonlinear activation function such as RELU or Sigmoid.

\subsection{Recurrence State Updater}

In the update stage, we first encode observed snapshot $G_{t}$ with the bi-flow encoder as $\varepsilon(G_{t}) \in \mathbb{R}^{n \times d_{\varepsilon}}$. Next, to learn the vector representation of time $t$ which can be fed into the neural network to capture both periodic and non-periodic patterns in graph sequence, we use the following TIME2VEC~\cite{kazemi2019time2vec} technique:
\begin{equation}
    \begin{split}
f_{T}(t)[r] = \left \{
\begin{array}{ll}
    w_{r}t+\varphi_{r},                    & \text{if } r=0 \\
    sin(w_{r}t+\varphi_{r}),     & \text{if } 1<r \leq d_{T} \\
\end{array}
\right.
    \end{split}
\end{equation}
where $f_{T}(t) \in \mathbb{R}^{d_{T}}$ is the representation for the timestep $t$. $w_{1},w_{2},...w_{d_{T}},\varphi_{1},\varphi_{2},...,\varphi_{d_{T}} \in \mathbb{R}$ are trainable parameters which are shared across different timesteps. Then, we concatenate the encoded node state $\varepsilon(v_{i,t})$, latent variable $z_{i,t}$, and time vector $f_{T}(t)$, and update the node hidden state with a GRU cell to obtain $h_{i,t}$, which denotes the newly updated hidden state of node $v_{i}$. 

\subsection{Joint Optimization Strategy}

To jointly learn the co-evolution pattern of graph structure and node attributes in a graph sequence with the variational recurrent model, we maximize the step-wise evidence lower bound (ELBO) for the log-likelihood of $p(G_{\leq T})$ as:

\begin{equation}
    \begin{split}
        \mathrm{log}p(G_{\leq T})& = -\mathrm{E}_{q_{\psi}(Z_{\leq T}|G_{\leq T})}\Big[\sum_{t=1}^{T}(\mathcal{L}_{prior}^{(t)}(q_{\psi}||p_{\phi})+\mathcal{L}_{rec}^{(t)}) \Big] \\
    \end{split}
\end{equation}
where $\mathcal{L}_{prior}^{(t)}$, $\mathcal{L}_{rec}^{(t)}$ are the prior regularization loss and reconstruction loss at timestep $t$, respectively.

Specifically, we minimize $\mathcal{L}_{prior}^{(t)}$ by making the posterior distribution stay close to the prior distribution. The discrepancy between the two distributions is measured by KL divergence:

\begin{equation}
    \mathcal{L}_{prior}^{(t)}(q_{\psi}||p_{\phi}) = \mathrm{KL}(q_{\psi}(Z_{\leq t}|Z_{< t},G_{\leq t})||p_{\phi}(Z_{\leq t}|Z_{< t},G_{< t}))
\end{equation}

$\mathcal{L}_{rec}^{(t)}$ reflects the step-wise reconstruction loss which can be naturally decomposed into structure reconstruction loss $\mathcal{L}_{struc}^{(t)}$ and attribute reconstruction loss $\mathcal{L}_{attr}^{(t)}$ as:

\begin{equation}
    \begin{split}
        \mathcal{L}_{rec}^{(t)} & = -\mathrm{log}p_{\varphi}(A_{t}, X_{t}|Z_{\leq t}, A_{<t}, X_{<t}) \\
            & = -\mathrm{log}p_{\varphi}(A_{t}|Z_{\leq t}, A_{<t}, X_{<t})-\mathrm{log}p_{\varphi}(X_{t}|Z_{\leq t}, A_{\leq t}, X_{<t}) \\
            & = \mathcal{L}_{struc}^{(t)} + \mathcal{L}_{attr}^{(t)} \\
    \end{split}
\end{equation}

For the structure reconstruction criterion $\mathcal{L}_{struc}^{(t)}$, we compute binary cross entropy (BCE) loss between $A_{t}$ and $\Tilde{A}_{t}$ as:
\begin{equation}
    \mathcal{L}_{struc}^{(t)} = -\frac{1}{|V|}\sum_{i=1}^{|V|}\sum_{j=1}^{|V|}[A_{ij,t}\mathrm{log}\Tilde{A}_{ij,t}+(1-A_{ij,t})\mathrm{log}(1-\Tilde{A}_{ij,t})]
\end{equation}

For attribute reconstruction loss $\mathcal{L}_{attr}^{(t)}$, instead of using mean square error (MSE), a common criterion in feature reconstruction~\cite{park2019symmetric,hou2022graphmae}, we adopt scaled cosine error (SCE), which can be formulated as:
\begin{equation}
    \mathcal{L}_{attr}^{(t)}=\frac{1}{|V|}\sum_{i=1}^{|V|}(1-\frac{x_{i,t}^{T}\Tilde{x}_{i,t}}{\Vert x_{i,t} \Vert \cdot \Vert \Tilde{x}_{i,t} \Vert})^{\alpha}, \alpha \geq 1,
\end{equation}
where $x_{i,t},\Tilde{x}_{i,t}$ denote original and generated node attributes for $v_{i}$ at timestep $t$, respectively. $\alpha$ is the scaling factor. Unlike MSE, this SCE is not sensitive to the impact of vector norm and dimensionality, improving the training stability of representation learning. Moreover, this criterion is selective enough to focus on those harder ones among imbalanced easy-and-hard samples. For easy samples with high-confidence predictions, their corresponding cosine errors are usually
smaller than 1 and decay faster to zero than hard samples when $\alpha \textgreater 1$. Thus, we can adaptively down-weight easy samples' loss in training and focus more on those hard ones.

\subsection{Generative Process}

\begin{algorithm}[t]
{
	\footnotesize  
	\caption{The VRDAG inference algorithm}
	\label{alg:Framework}
	\Input{$T$: time length, $p_{\phi}$: prior network, $\varepsilon$: bi-flow encoder, $p_{\varphi}$: attribute graph generator, $f_{T}$: TIME2Vec function, $\mathrm{GRU}$: GRU update cell}
	\Output{$\Tilde{G}$: Synthetic dynamic attributed graph}
        \vspace{0.5mm}
	\State{$(\Tilde{G},H_{0}) \gets (\varnothing,\textbf{0})$}

	\vspace{0.5mm}
	{
		\ForEach{$t \in [0,...,T-1]$}
		{
\Cmt{Temporal latent variables sampling}
			\State{$Z_{t+1} \sim p_{\phi}(H_{t})$}
   \vspace{0.5mm}
\Cmt{Attribute Graph Generation}
                \State{$\Tilde{A}_{t+1} \sim p_{\varphi}(A_{t+1}|Z_{\leq t},A_{<t+1},X_{<t+1})$}
\vspace{0.5mm}
                \State{$\Tilde{X}_{t+1} \sim p_{\varphi}(X_{t+1}|Z_{\leq t},A_{\leq t+1},X_{<t+1})$}
\vspace{0.5mm}
                \State{Get generated snapshot $\Tilde{G}_{t+1}(\Tilde{A}_{t+1},\Tilde{X}_{t+1})$}
                \vspace{0.5mm}
\Cmt{Hidden State Update}

                \State{$H_{t+1} = \mathrm{GRU}(\left[\varepsilon(\Tilde{G}_{t+1})||Z_{t+1}||f_{T}(t+1)\right],H_{t})$}
                \vspace{0.5mm}
			\State{$\Tilde{G} \gets \Tilde{G} \cup \{\Tilde{G}_{t+1}
    (A_{t+1},X_{t+1})\}$}
		}
	}
 \Return{$\Tilde{G}$};
}
\end{algorithm}

With the trained model, we can generate a new dynamic attributed graph $\Tilde{G}$ from scratch. The overall VRDAG inference algorithm is described in Algorithm~\ref{alg:Framework}. Before the generative process, we initialize an empty set to collect generated graph snapshots and initialize the hidden state $H_{0} \in \mathbb{R}^{N \times d_{h}}$ as a zero vector matrix (Line 1). During recurrent generation, we sample the new latent node representation $Z_{t+1}$ from the prior distribution $p_{\phi}(H_{t})$ that is conditioned on historical graph information $H_{t}$ at each step (Line 3). Then, we generate the adjacency matrix of the graph at step $t+1$ using the trained MixBernoulli sampler (Line 4). Subsequently, we synthesize node attributes with the attribute decoder based on the generated topology $A_{t+1}$ and historical graph evolution information encoded in $Z_{t}$ and $H_{t}$ (Line 5). After obtaining the newly generated snapshot (Line 6), the model would update the hidden node embeddings. It first vectorizes the time $t+1$ using $f_{T}$ as $f_{T}(t+1)$ and encodes the previously generated $\Tilde{G}_{t+1}$ with the bi-flow encoder $\varepsilon$ as $\varepsilon(\Tilde{G}_{t+1})$. Then, the hidden states are updated to capture the time dependency in the graph sequence (Line 7). At the end of each generation step, we append the newly generated snapshot into the dynamic graph set (Line 8). Finally, after the recurrent generation described above, we return the synthetic dynamic attributed graph $\Tilde{G}$ with a given time length $T$ (Line 10).

\subsection{Complexity Analysis}

In this subsection, we provide a detailed complexity analysis of model parameters, model training, and model inference. For simplicity, we use $d$ to denote hidden units of deep neural networks utilized in our architecture. $\theta$ denotes model parameters. $L_{m}$ denotes the layer number of MLP used in our bi-flow encoder and $L$ is the number of GNN layers. 
$r$ is the number of sampled neighbors per node. $Q$ denotes the number of negative samples used in the structure reconstruction task. $S$ represents the number of training epochs and $l^{\prime}$ denotes the length of random walk. $M$ denotes the number of temporal edges. $C$ represents the number of components in the mixture model of TIGGER~\cite{gupta2022tigger}.

\myparagraph{Model complexity} For model complexity, we analyze each learnable module including bi-flow encoder ($\mathcal{O}(L \times L_{m} \times d^{2}+Fd)$), the prior/posterior distribution ($\mathcal{O}(d^{2})$), MixBernoulli Sampler ($\mathcal{O}(d^{2}+Kd)$), attribute decoder ($\mathcal{O}(d^{2})$), learnable time vector ($\mathcal{O}(d_{T})$), and GRU cell ($\mathcal{O}(d^{2}+d\times d_{T})$). The overall model complexity $\mathcal{O}(|\theta|)$ can be simplified to $\mathcal{O}(L \times L_{m} \times d^{2}+Fd+d\times d_{T})$.

\myparagraph{Training complexity} For model training, the prior and posterior distribution sampling both take a complexity of $\mathcal{O}(Nd^{2})$. The complexity of bi-flow graph encoding has a complexity of $\mathcal{O}(L(Ndr+L_{m}Nd^{2}))$. Topology decoding and attribute decoding cost $\mathcal{O}(N^{2}d)$ and $\mathcal{O}(L(Ndr+Nd^{2}))$, respectively. The complexity of time vectorization and GRU updates are $\mathcal{O}(d_{T})$ and $\mathcal{O}(N(d^{2}+d\times d_{T}))$. Calculating $\mathcal{L}_{prior}^{(t)},\mathcal{L}_{struc}^{(t)}$ and $\mathcal{L}_{attr}^{(t)}$ take a complexity of $\mathcal{O}(Nd)$, $\mathcal{O}(Nr+NQ)$, and $\mathcal{O}(Nd)$, respectively. The complexity of the backward process is $\mathcal{O}(|\theta|)$. Considering the length of time steps $T$ and training epochs $S$, the final training complexity is $\mathcal{O}(TS(N^{2}d+L(Ndr+L_{m}Nd^{2})))$.

\myparagraph{Inference complexity} For complexity analysis of model inference, the computation of posterior sampling, loss function, and backward propagation in the training stage are removed. Then, we obtain the inference complexity as $\mathcal{O}(T(N^{2}d+L(Ndr+L_{m}Nd^{2})))$. The most efficient temporal random walk-based generator TIGGER~\cite{gupta2022tigger} has a complexity of $\mathcal{O}(M \times l^{\prime} \times (N+C))$. As $M \gg N$ in many real-world dynamic graphs, our VRDAG, which eliminates the need for extensive path sampling, can greatly facilitate generation efficiency. Besides, parallel computation on GPU further accelerates our model compared to the sampling-based methods.

\subsection{Extension}
As stated in Section~\ref{sec:pre}, we consider all unique nodes $V=\cup_{t=1}^{T}V_{t}$ in the dynamic graph $G$ and model the structural evolution as changes in $E_{t}$. To extend our architecture to support flexible node addition and deletion, we discuss the potential solutions: (1) For node deletion, we consider each node $v_{i}$ and maintain a counter to track its consecutive time steps of isolation from other nodes in the graph. If this count reaches the predefined threshold $T_{del}$, we remove its hidden node state in the sequential generation, ensuring it no longer exists in the new snapshot generation. (2) For node addition, we propose to train a predictor to estimate the newly added node number, $N_{add}$, based on the hidden graph state $\Bar{h}_{t}=\frac{1}{N_{t}}\sum_{i=1}^{N_{t}}h_{i,t}$, which is calculated as the average hidden node state. Additionally, we parameterize a distribution $p_{\omega}$ to sample the initial hidden states $H_{t}^{\prime} \in \mathbb{R}^{N_{add} \times d_{h}}$ for these added nodes. The distribution is conditioned on the time vector $f_{T}(t)$ and the hidden graph state $\Bar{h}_{t}$. Then, we stack the original hidden node state $H_{t}$ with the sampled $H_{t}^{\prime}$ and feed them into our attributed graph decoder to generate the new snapshot.

\section{Experiments}
\label{sec:exp}

\subsection{Experiment Settings}

\begin{table*}[ht]
\caption{The statistics of the datasets and the overall experimental results for evaluating graph structure generation in terms of different network properties are presented. Here, N and M denote the number of nodes and temporal edges, respectively, and X represents the number of node attributes.}
\vspace{-5px}
\label{tab:comp}
\centering
\renewcommand\arraystretch{1.2}
\begin{tabular}{cccccccccccccc}
\hline
\textbf{Dataset} & \textbf{N}=$|V|$ & \textbf{M}=$|E|$ & \textbf{X} & \textbf{T} & \textbf{Method} & \begin{tabular}[c]{@{}c@{}} \textbf{In-deg} \\ \textbf{dist} \end{tabular}& \begin{tabular}[c]{@{}c@{}} \textbf{Out-deg} \\ \textbf{dist} \end{tabular} & \begin{tabular}[c]{@{}c@{}} \textbf{Clus} \\ \textbf{dist} \end{tabular} & \begin{tabular}[c]{@{}c@{}} \textbf{In-PLE}  \end{tabular} & \begin{tabular}[c]{@{}c@{}} \textbf{Out-PLE}  \end{tabular} &\begin{tabular}[c]{@{}l@{}} \textbf{Wedge} \\ \textbf{count} \end{tabular} & \textbf{NC} & \textbf{LCC} \\ \hline

\multirow{7}{*}{Emails-DNC} & \multirow{7}{*}{1,891} & \multirow{7}{*}{39,264}    & \multirow{7}{*}{2}        & \multirow{7}{*}{14} & GRAN & 0.0249 & 0.0262 & 0.2116 & 0.1491 & 0.2792 & 0.6165 & 0.9095 & 0.9125 \\
                  &                   &                                                        &        &            & GenCAT &  0.0083 & 0.0076 & \textbf{0.0056} & 0.3559 & 0.3321 & 3.7134 & 0.8875 & 0.1889 \\
                  &                   &                                                        &         &           & TagGen &  0.0052 & 0.0041 & 0.0086 & 0.0719 & 0.0803 & 0.0671 & 0.9194 & \textbf{0.0943} \\
                  &                   &                                                        &         &           & Dymond &  0.0071 & 0.0093 & 0.0411 & 0.1465 & 0.1585 & 1.2573 & 0.1388 & 0.5807 \\
                  &                   &                                                        &        &            & TGGAN & 0.0082 & 0.0075 & 0.0636	& 0.1013 & 0.1388 & 0.1253 & 0.2971	& 0.1584 \\
                  &                   &                                                        &        &            & TIGGER &  0.0173 & 0.0161 & 0.0066 & 0.1335 & 0.2243 & 4.2082 & \textbf{0.1063} & 0.1723 \\
                  &                   &                                                        &        &            & VRDAG & 
                  \textbf{0.0026} & \textbf{0.0038} & 0.0312 & \textbf{0.0297} & \textbf{0.0486} & \textbf{0.0389} & 0.1520 & 0.5395 \\ \hline
\multirow{6}{*}{Bitcoin-Alpha} & \multirow{6}{*}{3,783} & \multirow{6}{*}{24,186}    & \multirow{6}{*}{1}                                   & \multirow{6}{*}{37} & GRAN & 0.5512 & 0.5393 & 0.4626 & 0.2387 & 0.1910 & 0.9980 & 0.2693 & 0.9367 \\
                  &                   &                                                        &       &            & GenCAT & 0.2229 & 0.1965 & 0.0096 & 0.3177  & 0.3428 & 9.1801 & 0.6250 & \textbf{0.0044}  \\
                  &                   &                                                        &       &           & TagGen & 0.0079 & 0.0146 & 0.0102 & 0.1898 & 0.1933 & 0.2694 & 0.8922 & 0.0184 \\
                   &                   &                                                        &      &             & TGGAN & 0.0105 & 0.0112	& 0.0247 & 0.0504 & 0.0473	& 0.3126	& 0.6642 & 0.0093 \\
                  &                   &                                                        &       &            & TIGGER & 0.0069 & 0.0075 & 0.0072 & 0.0021 & 0.0296 & \textbf{0.0824} & 0.7683 & 0.0051 \\
                  &                   &                                                        &       &            & VRDAG & \textbf{0.0057} & \textbf{0.0029} & \textbf{0.0066} & \textbf{0.0014} & \textbf{0.0178} & 0.4327 &  \textbf{0.2485} & 0.0442 \\ \hline
\multirow{6}{*}{Wiki-Vote} & \multirow{6}{*}{7,115} & \multirow{6}{*}{103,689}    & \multirow{6}{*}{1}                                   & \multirow{6}{*}{43} & GRAN & 0.8726 & 0.9414 & 0.3856 & 0.2611 & 0.6961 & 0.9988 & 0.9418 & 0.8991 \\
                  &                   &                                                        &        &           & GenCAT & 0.0198 & 0.0264 & 0.2953 & 0.2130 & 0.9794 & 1.6067 & 0.7836 & 0.1355 \\
                  &                   &                                                        &         &          & TagGen & 0.0196 & 0.0232 & \textbf{0.0119} & \textbf{0.0944} & 0.0779 & 0.2169 & 1.6582 & \textbf{0.0072} \\
                     &                   &                                                        &      &             & TGGAN & 0.0325 & 0.0277 & 0.0228 & 0.0709 & 0.0983 & 0.2085 & 0.9274 & 0.0104 \\
                  &                   &                                                        &         &          & TIGGER & 0.0411 & 0.0549 & 0.0188 & 0.1273 & 0.2168 & 2.0239 & 2.8787 & 0.0466 \\
                  &                   &                                                        &       &            & VRDAG & \textbf{0.0061} & \textbf{0.0081} & 0.0132 & 0.1845 & \textbf{0.0761} & \textbf{0.1477} & \textbf{0.7479} & 0.5661 \\ \hline
\multirow{6}{*}{Guarantee} & \multirow{6}{*}{5530} & \multirow{6}{*}{6,169}       & \multirow{6}{*}{2}                               & \multirow{6}{*}{15} & GRAN & 0.0052 & 0.0029 & 0.0085 & 0.9128 & 0.4364 & 0.8760 & 0.9891 & 12.4904 \\
                  &                   &                                                        &       &            & GenCAT & 0.0018 & 0.0049 & 0.0318 & 0.9432 & 0.2602 & 4.0482 & 0.8387 & 54.892  \\
                  &                   &                                                        &       &            & TagGen & 0.0188 & 0.0275 & 0.0670 & 0.6455 & 0.2304 & 26.629 & 0.1728 & 72.594 \\
                    &                   &                                                        &      &             & TGGAN & 0.0057 & 0.0065	& 0.0421 & 0.5136 & 0.3735 & 3.4952 & \textbf{0.1075}	& 4.6509 \\
                  &                   &                                                        &         &          & TIGGER & 0.0019 & 0.0043 & 0.0015 & 0.8673 & 0.2769 & 0.8738 & 0.1879 & 1.2869 \\
                  &                   &                                                        &        &           & VRDAG & \textbf{0.0013} & \textbf{0.0015} & \textbf{0.0009} & \textbf{0.4915} & \textbf{0.1834} & \textbf{0.7649} & 0.1604 & \textbf{0.8784} \\ \hline
\multirow{4}{*}{Brain} & \multirow{4}{*}{5,000} & \multirow{4}{*}{529,093}             & \multirow{4}{*}{20}                          & \multirow{4}{*}{12} & TagGen & 0.0135 & 0.0259	& 0.0551 & 0.1446 & 0.1532 & 0.3992 & 0.3285	& 0.2392 \\
                    &                   &                                                        &      &             & TGGAN & 0.0177 & 0.0162& 0.0648 & 0.1638 & 0.1402 & 0.5260 & 0.7330 & 0.5411 \\
                  &                   &                                                        &        &           & TIGGER & 0.0053 & 0.0059 &	\textbf{0.0322} & 0.1362 & 0.2587 & 0.5518 & \textbf{0.1939} & \textbf{0.0802} \\
                  &                   &                                                        &        &           & VRDAG & \textbf{0.0025} & \textbf{0.0041} &	0.0475 & \textbf{0.0873} & \textbf{0.1334} & \textbf{0.2623} & 0.2258 & 0.3648 \\ \hline
\multirow{4}{*}{GDELT} & \multirow{4}{*}{5,037} & \multirow{4}{*}{566,735}       & \multirow{4}{*}{10}                               & \multirow{4}{*}{18} & TagGen & 0.0038 & 0.0035	& \textbf{0.0141} & 0.1683 & 0.1476 & \textbf{0.7726} & 1.1441 & \textbf{0.0581} \\
                    &                   &                                                        &        &           & TGGAN & 0.0074 & 0.0062& 0.0356 & 0.3840 & 0.3066 & 1.3965 & 2.0256 & 0.0817 \\
                  &                   &                                                        &          &         & TIGGER & 0.0117 & 0.0091	& 0.0676 & 0.2012 & 0.1938 & 1.8698 & 1.4718 & 0.4528 \\
                  &                   &                                                        &          &         & VRDAG & \textbf{0.0021} & \textbf{0.0017} & 0.0283 & \textbf{0.0524} & \textbf{0.0767} & 0.9121 & \textbf{0.7574} &0.3089 \\ \hline
\end{tabular}
\end{table*}

\subsubsection{Dataset and Baselines}
\label{data_base}

We use six dynamic graph datasets including five publicly available graphs (Emails-DNC (Email)~\cite{nr}, Bitcoin-Alpha (Bitcoin)~\cite{kumar2016edge}, Wiki-Vote (Wiki)~\cite{leskovec2010predicting}, Brain~\cite{xu2019spatio}, and GDELT~\cite{zhou2022tgl}) and a guaranteed-loan network (Guarantee) collected from a major commercial bank. The detailed dataset descriptions can be found in Appendix A-A~\cite{onlineapp}. For compared baselines, we adopt four dynamic graph generators (i.e., TagGen~\cite{zhou2020data}, Dymond~\cite{zeno2021dymond}, TGGAN~\cite{zhang2021tg}, and TIGGER~\cite{gupta2022tigger}) and two powerful static graph generators including GRAN~\cite{liao2019efficient} and GenCAT~\cite{maekawa2023gencat}. GenCAT is the latest state-of-the-art attributed graph generator. Detailed descriptions for the compared methods can be found in Appendix A-B~\cite{onlineapp}. We have provided detailed dataset statistics in TABLE~\ref{tab:comp}.

\subsubsection{Evaluation Metrics}
\label{metric}

The evaluation metrics are categorized into graph structure metrics, node attribute metrics, and difference metrics. We leave the detailed node attribute metrics in Appendix A-C~\cite{onlineapp}.

\myparagraph{Graph structure metrics} To evaluate structure generation performance, we consider eight widely used network properties~\cite{xiang2022efficient,kocayusufoglu2022flowgen,gupta2022tigger}, including in/out-degree distribution (In/Out-deg dist), clustering coefficients distribution (Clus dist), power-law exponent of in/out-degree distribution (In/Out-PLE), wedge count, number of components (NC), and size of the largest connected components (LCC). For the degree distribution and clustering coefficient distribution,
we compute the Maximum Mean Discrepancy (MMD)~\cite{xiang2022efficient} to quantify the distribution difference between the original snapshot and the synthetic snapshot for each unique timestep. The average distribution discrepancy across the graph sequence is then reported. Regarding other node or graph-level metrics, we follow the approach described in~\cite{zhou2020data} and extend them to the dynamic setting by measuring the average discrepancy in percentage as:
\begin{equation}
    \mathcal{M}_{avg}(G,\Tilde{G}) = Mean_{t=1:T}\frac{|\mathcal{M}(G_{t})-\mathcal{M}(\Tilde{G_{t}})|}{\mathcal{M}(G_{t})}
\end{equation}
where $\mathcal{M}(\cdot)$ denotes a specific metric function. 
 

\myparagraph{Difference metrics} To measure the difference between two consecutive snapshots, we compute the differences for each pair of nodes with the same ID in the neighboring snapshots using three structural properties (degree, clustering coefficient, and coreness) and two attribute metrics (MAE and RMSE), and report the average result of each metric. Specifically, consider the consecutive snapshots $G_{t}$ and $G_{t+1}$, the node with the same ID on two graphs is denoted as $v_{i,t}$ and $v_{i,t+1}$, respectively. For structure difference, we let the structural property (e.g., degree) for the node as $P_{s}$, and the topology difference $D_{s}$ can be measured as: 
\begin{equation}
    D_{s}(G_{t},G_{t+1}) = \frac{1}{N}\sum_{i=1}^{N}|P_{s}(v_{i,t})-P_{s}(v_{i,t+1})|
\end{equation}
For attribute discrepancy, we denote $x_{i,t},x_{i,t+1}$ as attribute data of $v_{i}$ in the two consecutive snapshots. The MAE and RMSE can be calculated as:
\begin{equation}
    \begin{aligned}
        \mathrm{MAE}(X_{t},X_{t+1}) & = \frac{1}{N}\sum_{i=1}^{N}|x_{i,t}-x_{i,t+1}| \\
        \mathrm{RMSE}(X_{t},X_{t+1}) & = \sqrt{\frac{1}{N}\sum_{i=1}^{N}(x_{i,t}-x_{i,t+1})^{2}} \\
    \end{aligned}
\end{equation}
For multi-dimensional attributes, we average the discrepancy along the attribute dimension in implementation.

More implementation details of our experiments can be found in Appendix A-D~\cite{onlineapp}

\subsection{Evaluation of Generated Graphs}

In this section, we evaluate the performance of our model in structure generation and node attribute generation. 

\begin{figure}[t!]
	\centering
	\subfloat[JSD]{
		\includegraphics[width=0.45\linewidth]{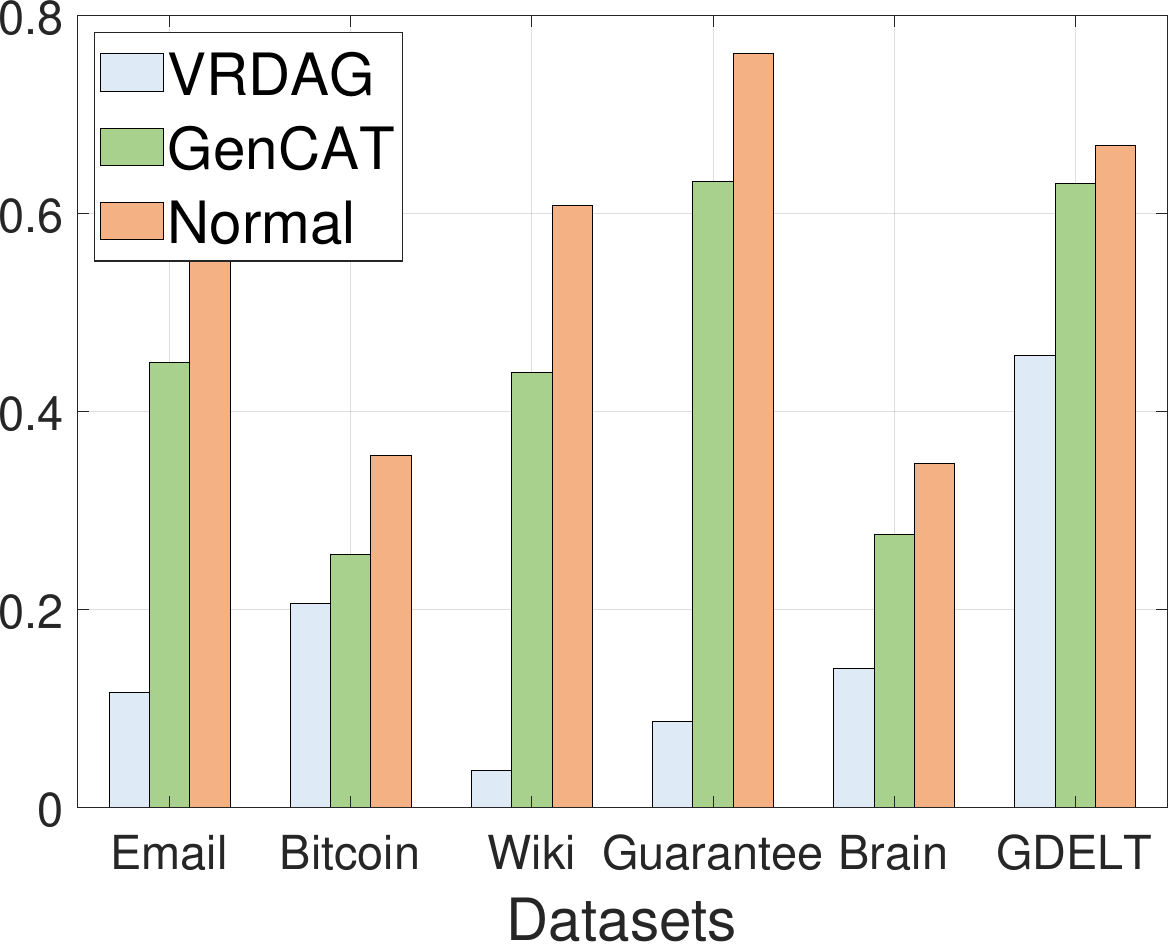}}
	\subfloat[EMD]{
		\includegraphics[width=0.45\linewidth]{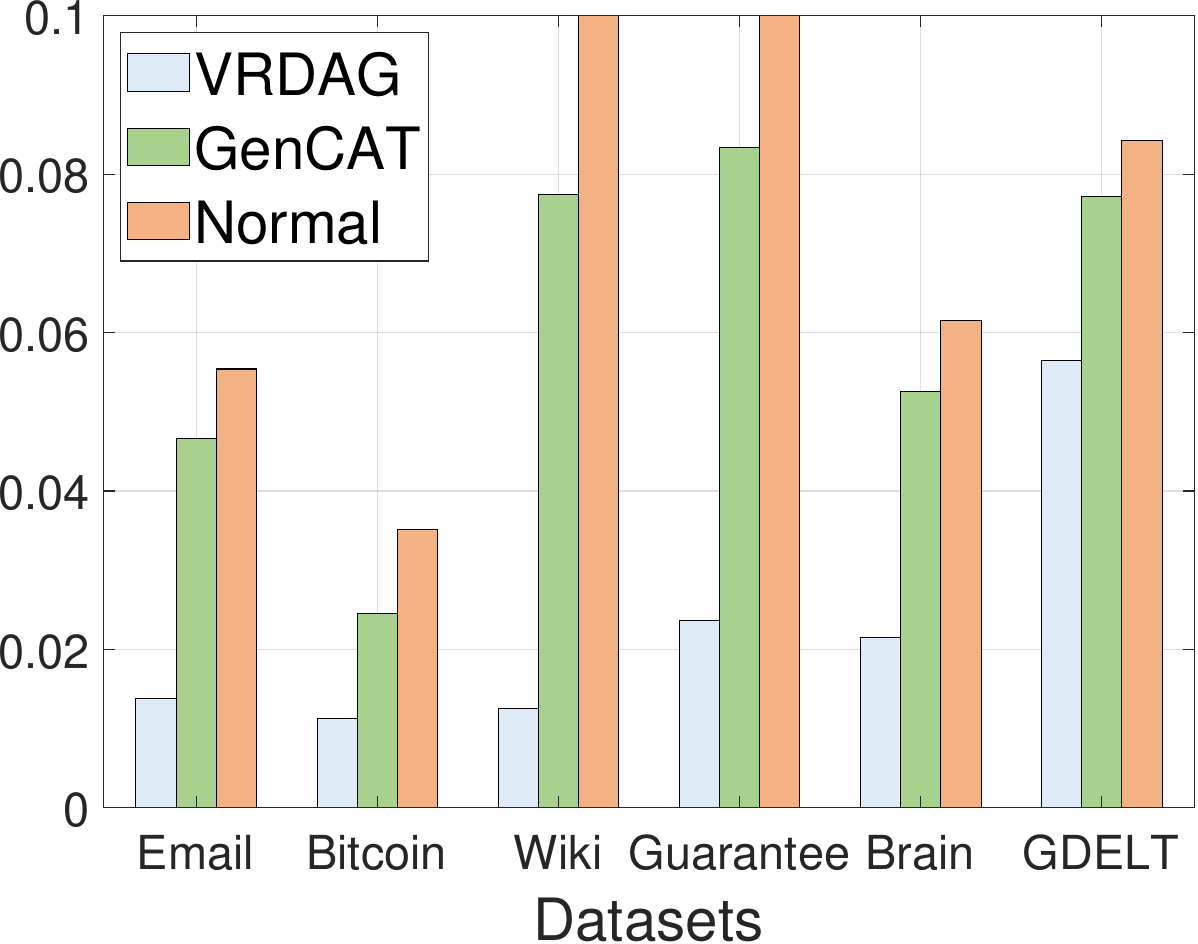}}
        \caption{Evaluation results for attribute distribution. Normal denotes the normal distribution where the mean and variance values are estimated from the ground-truth data. For the Earth Mover's Distance, we cut off
high values for better visibility.}
        \label{fig:attr}
\end{figure}

\myparagraph{Network structure generation performance}
TABLE~\ref{tab:comp} presents the performance of VRDAG compared to recently proposed strong graph generative models, evaluated using eight graph structure metrics on six real-world datasets. Note that the motif-based dynamic method Dymond can only be executed on the Email dataset, the smallest dataset, due to its requirement for the storage of millions of motif structures across time. Overall, VRDAG and three models based on temporal random walk consistently achieve the best results across most graph structure metrics. In contrast, the static methods GRAN and GenCAT perform poorly as they cannot learn the evolution patterns in graph sequence data. Although GenCAT outperforms GRAN in terms of distribution discrepancy metrics, it fails to generate representative substructures found in the original graphs. This is because GenCAT can only model a specific structural distribution, unlike deep models that can effectively learn complex structural knowledge. TagGen, TGGAN, and TIGGER are the three most highly competitive models. They employ temporal random walks to capture joint distribution between graph topology and time properties, leading to superior performance in dynamic graph generation compared to previous methods. Our VRDAG demonstrates superior structure generation performance than the three learning-based dynamic methods. Specifically, VRDAG achieves the best results in 6 out of the 8 metrics on Bitcoin dataset and more than half of the best results on the Wiki dataset. On large datasets Brain and GDELT, which also contain more attribute information, VRDAG significantly improves topology synthesis. This suggests that the auxiliary node attribute knowledge may help capture hidden links between nodes. For instance, the generated dynamic graph matches the in/out-degree distribution of the original data, with MMD values not exceeding 4.1$ \times 10^{-3}$ and 2.1$ \times 10^{-3}$ on Brain and GDELT, respectively. The model also achieves the lowest discrepancy value in PLE metrics on the above two datasets. These findings indicate that the model effectively captures network
message flow distribution, which is significant in realistic graph generation. 
Moreover, VRDAG remains superior in other network properties such as wedge count and number of components, which could reflect specific connection relationships in the ground-truth data. For example, on the GDELT dataset, the NC error of VRDAG is 1.26 lower than that of TGGAN.
We also notice that VRDAG does not offer superior performance on certain metrics and datasets, especially denser graphs. This may be due to our graph-based VAE architecture, which tends to generate highly similar or ‘smoothed’ features in dense graphs, resulting in a lack of details in the generated graphs.
Additionally, the sampling process may introduce noise and local inconsistencies. For the mediocre performance in the LCC metric, we analyze that this is because our VAE-based framework compresses the graph into a low-dimensional node latent space, which focuses more on pairwise relationships between nodes and results in a lack of long-range connectivity essential for generating accurate LCC, unlike temporal random-walk methods that ensure stronger adjacency-based connectivity.

\begin{table}
\caption{The mean absolute error across Spearman's correlation coefficients of attributes on two real-world datasets}
\label{tab:attrsp}
\tabcolsep 10pt
\centering
\begin{tabular}{cccc}
\toprule
    Dataset   & Normal & GenCAT & VRDAG \\ 
  \midrule
    Email     &   0.9031     &   0.7309     & \textbf{0.1389}      \\
    Guarantee &    1.1713    &   1.0018    &  \textbf{0.2440}     \\ 
  \bottomrule
\end{tabular}
\end{table}

\myparagraph{Node attribute generation performance}
Fig.~\ref{fig:attr} illustrates the average JSD and EMD between the synthetic and original dynamic graphs using three node attribute generation methods. Since there are no existing dynamic attributed graph generative models, we compare VRDAG with a recently proposed static node attribute generation method, GenCAT, and a random generation method based on the normal distribution (Normal). 
Based on the experimental results, our VRDAG consistently outperforms these two methods in preserving the distribution of node attributes. For instance, the JSD values corresponding to our VRDAG model are less than 0.1 on the Wiki and Guarantee datasets, while GenCAT exhibits distances that are at least 5 times larger than VRDAG. Besides, VRDAG achieves at least 0.135 lower JSD values on Brain and GDELT datasets compared to GenCAT. This validates the superiority of our framework in generating multiple structure-independent node attributes.
Additionally, VRDAG outperforms two baselines in terms of EMD, achieving a discrepancy of below $2.5 \times 10^{-2}$ on five out of the six datasets. On GDELT, the EMD value achieved by VRDAG is 26.8\% lower than that of GenCAT. We also observe that the simple normal distribution is inadequate for approximating the unknown and complex node attribute distributions. The evaluation results clearly indicate that our data-driven method can effectively maintain attribute distribution consistency with the original dynamic graph, whereas traditional static methods fall short in capturing the temporal properties of node attributes and fail to model the underlying complex distribution. Moreover, as shown in Table~\ref{tab:attrsp}, the errors in correlation coefficients achieved by VRDAG are 80.9\% and 75.6\% smaller than those of the state-of-the-art GenCAT on the Email and Guarantee datasets, respectively. This indicates that our model can preserve attribute correlations, unlike Normal and GenCAT which treat each attribute as an independent variable.

\textit{Our ablation study and parameter analysis can be found in Appendix A-E~\cite{onlineapp}, and Appendix A-F~\cite{onlineapp}, respectively.}

\begin{figure} [t]
	\centering
	\subfloat[Email]{
		\includegraphics[width=0.32\linewidth]{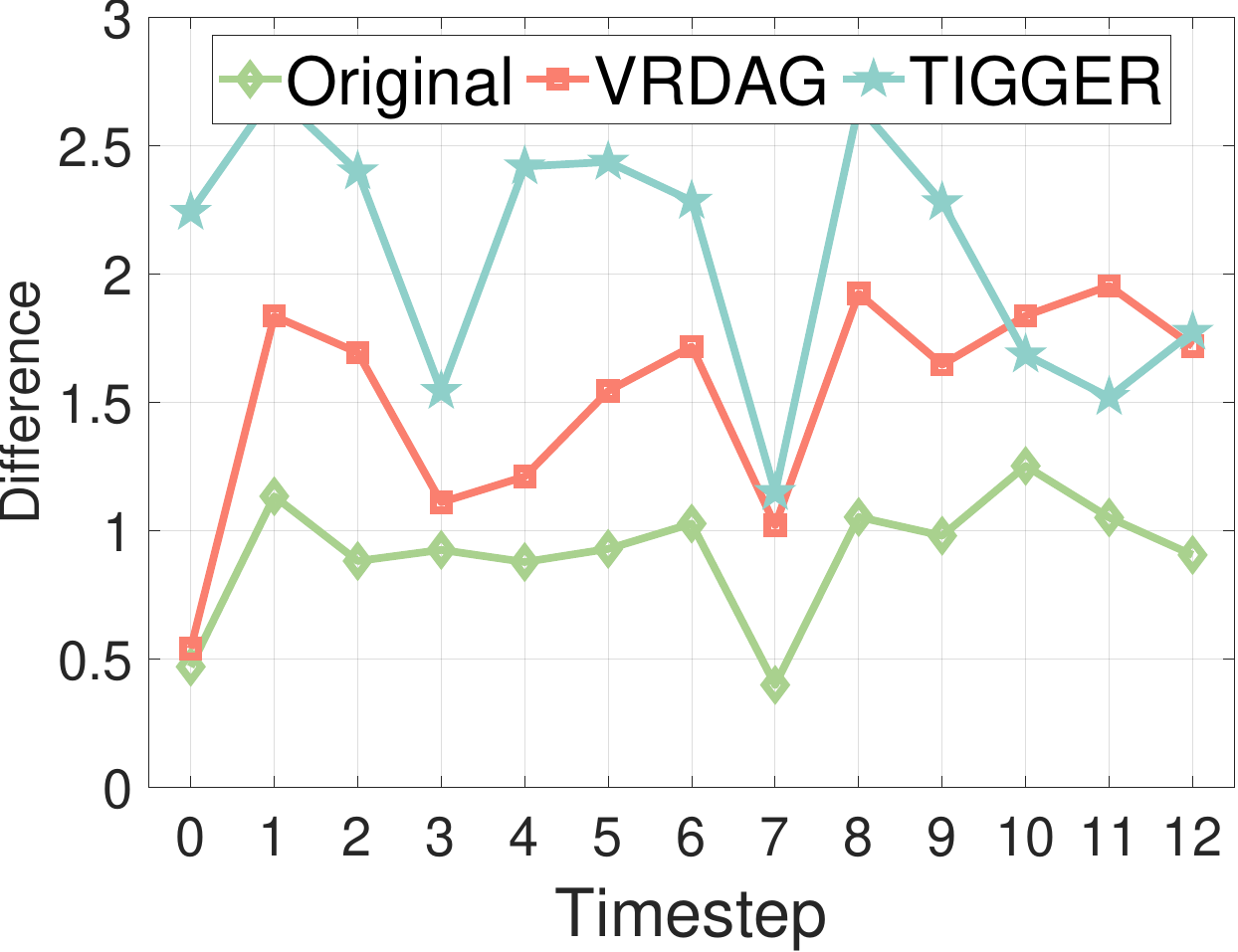}}
	\subfloat[Wiki]{
		\includegraphics[width=0.31\linewidth]{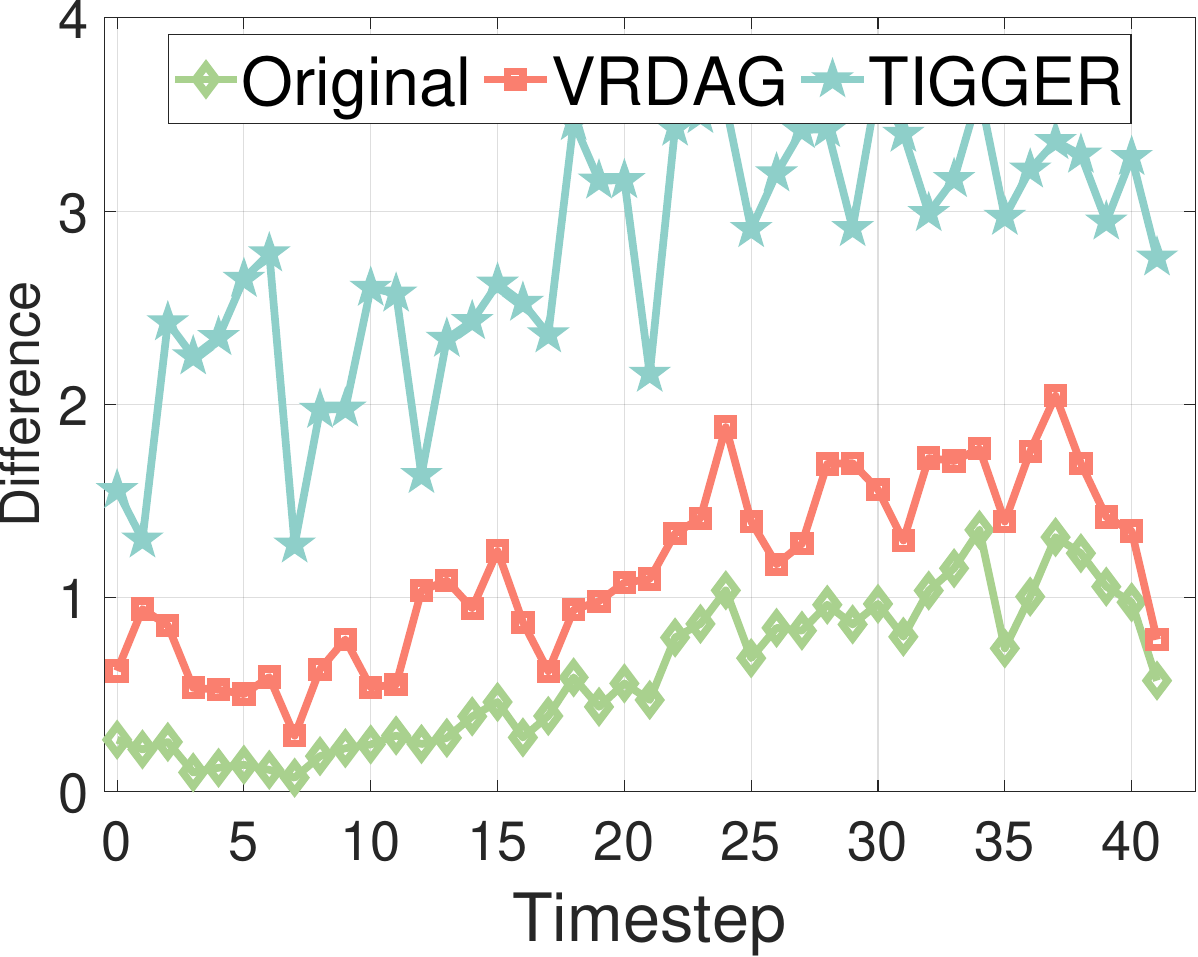}}
        \subfloat[GDELT]{
		\includegraphics[width=0.32\linewidth]{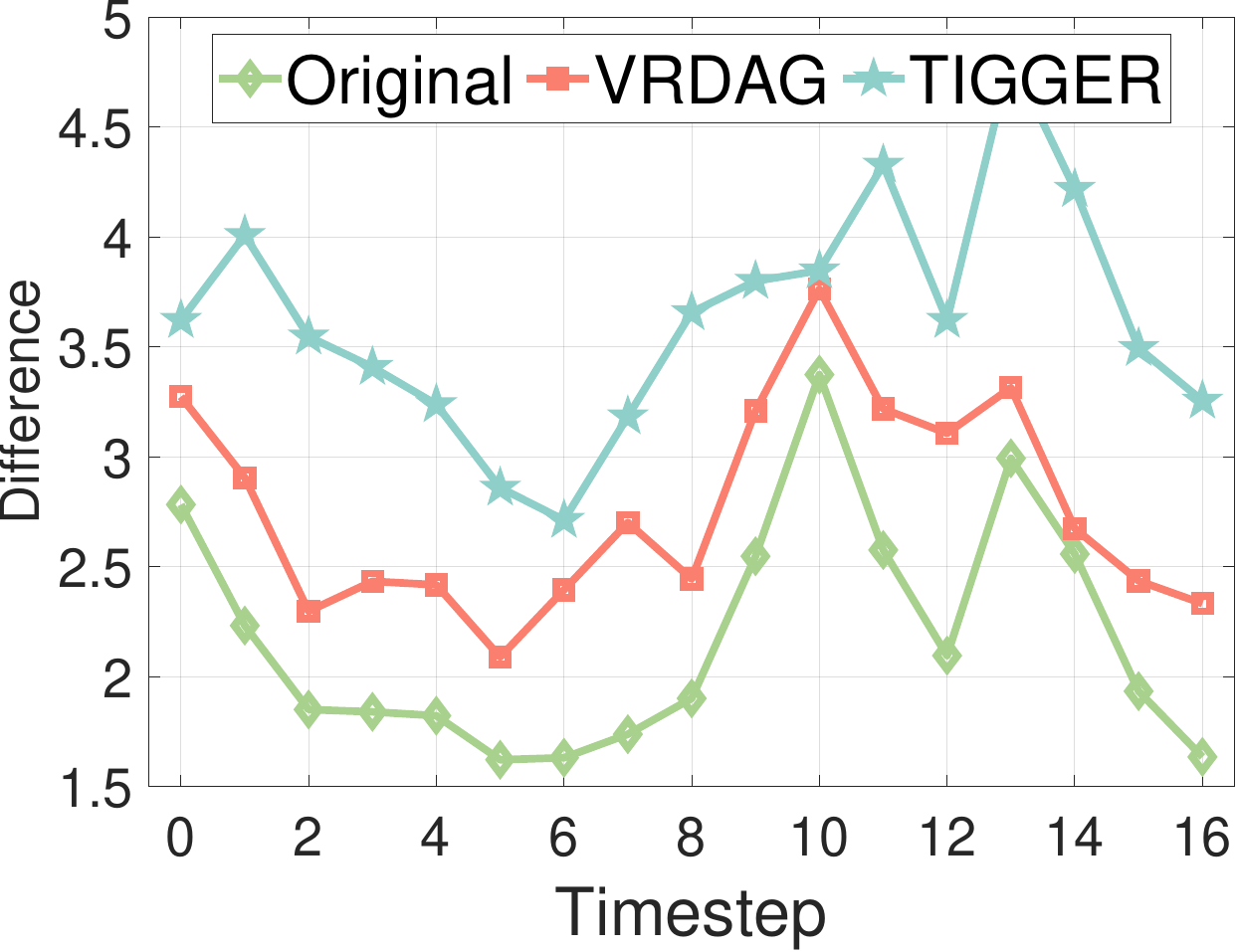}} 
        \caption{Temporal structure difference in degree}
        \label{fig:deg}
\end{figure}

\begin{figure}[t]
	\centering
	\subfloat[Email]{
		\includegraphics[width=0.32\linewidth]{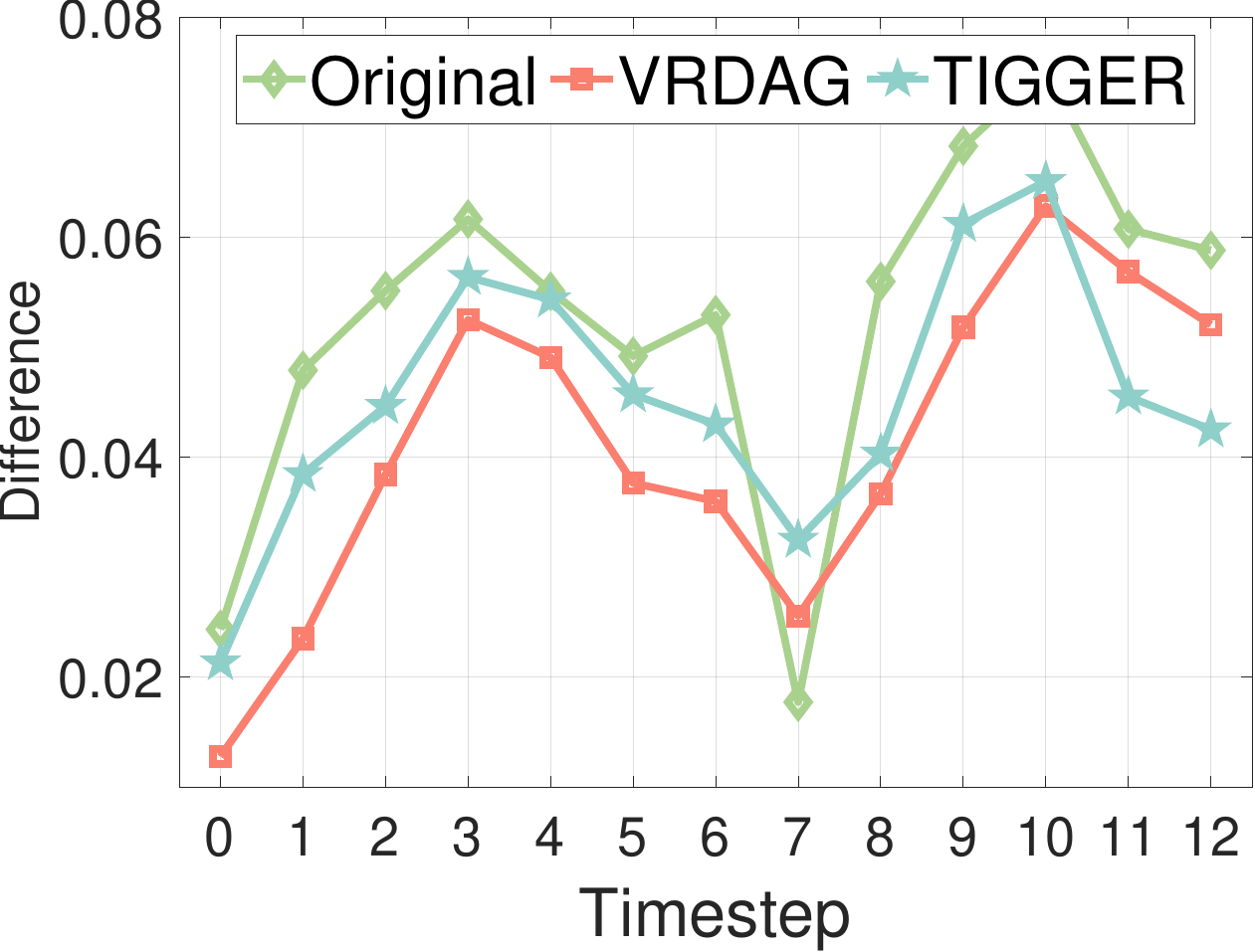}}
	\subfloat[Wiki]{
		\includegraphics[width=0.32\linewidth]{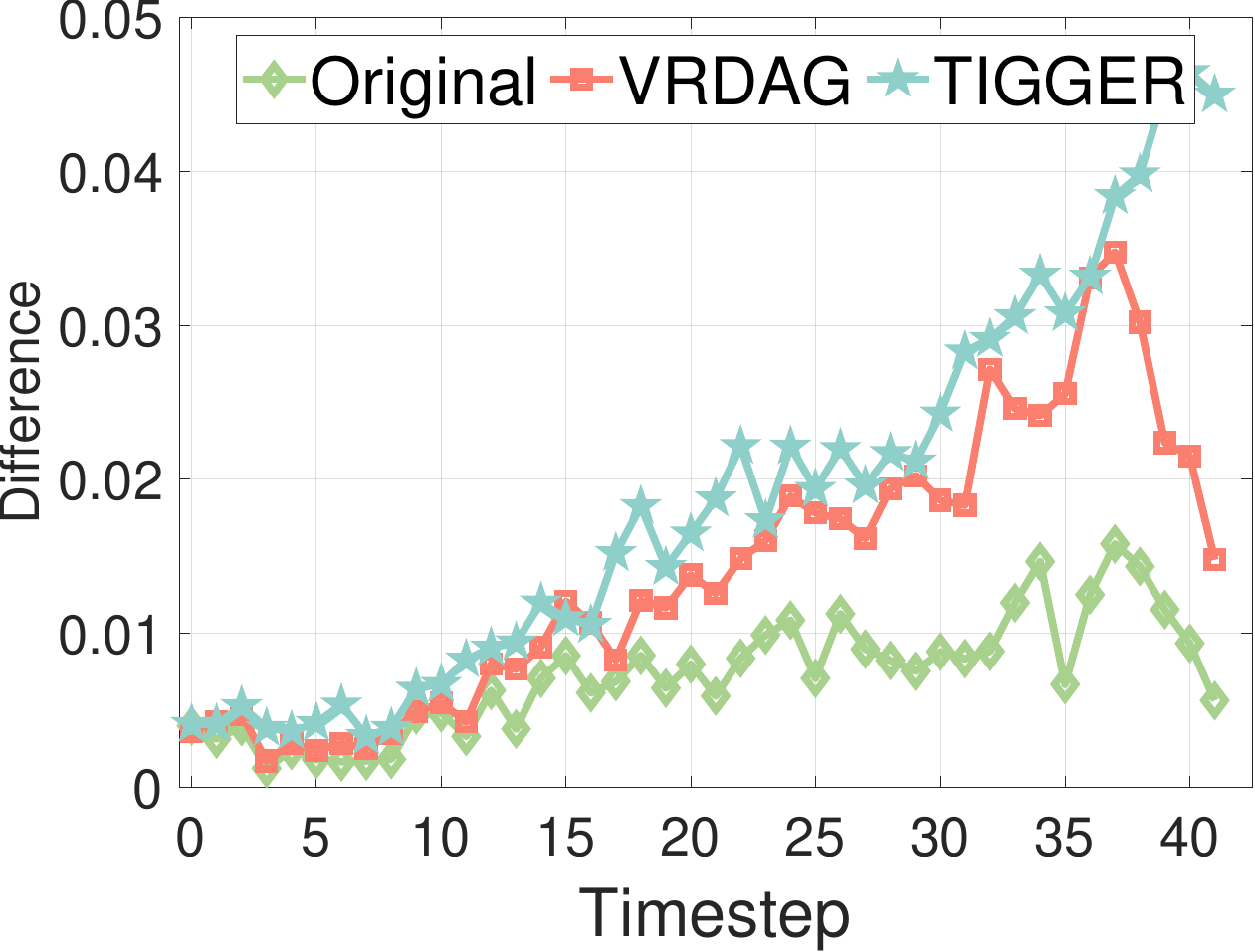}}
        \subfloat[GDELT]{
		\includegraphics[width=0.32\linewidth]{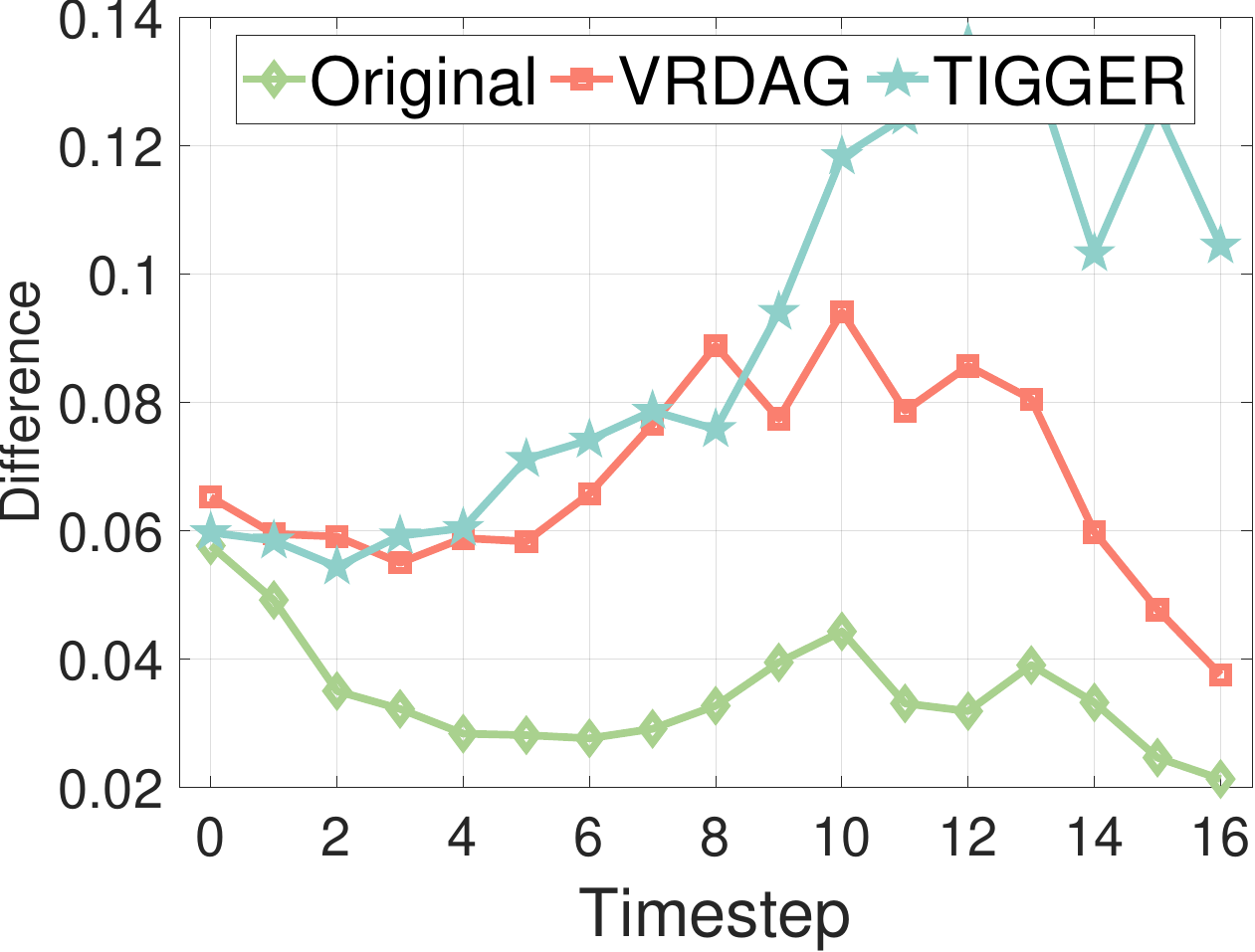}} 
        \caption{Temporal structure difference in clustering coefficient}
        \label{fig:clus}
\end{figure}

\begin{figure}[t]
	\centering
	\subfloat[Email]{
		\includegraphics[width=0.32\linewidth]{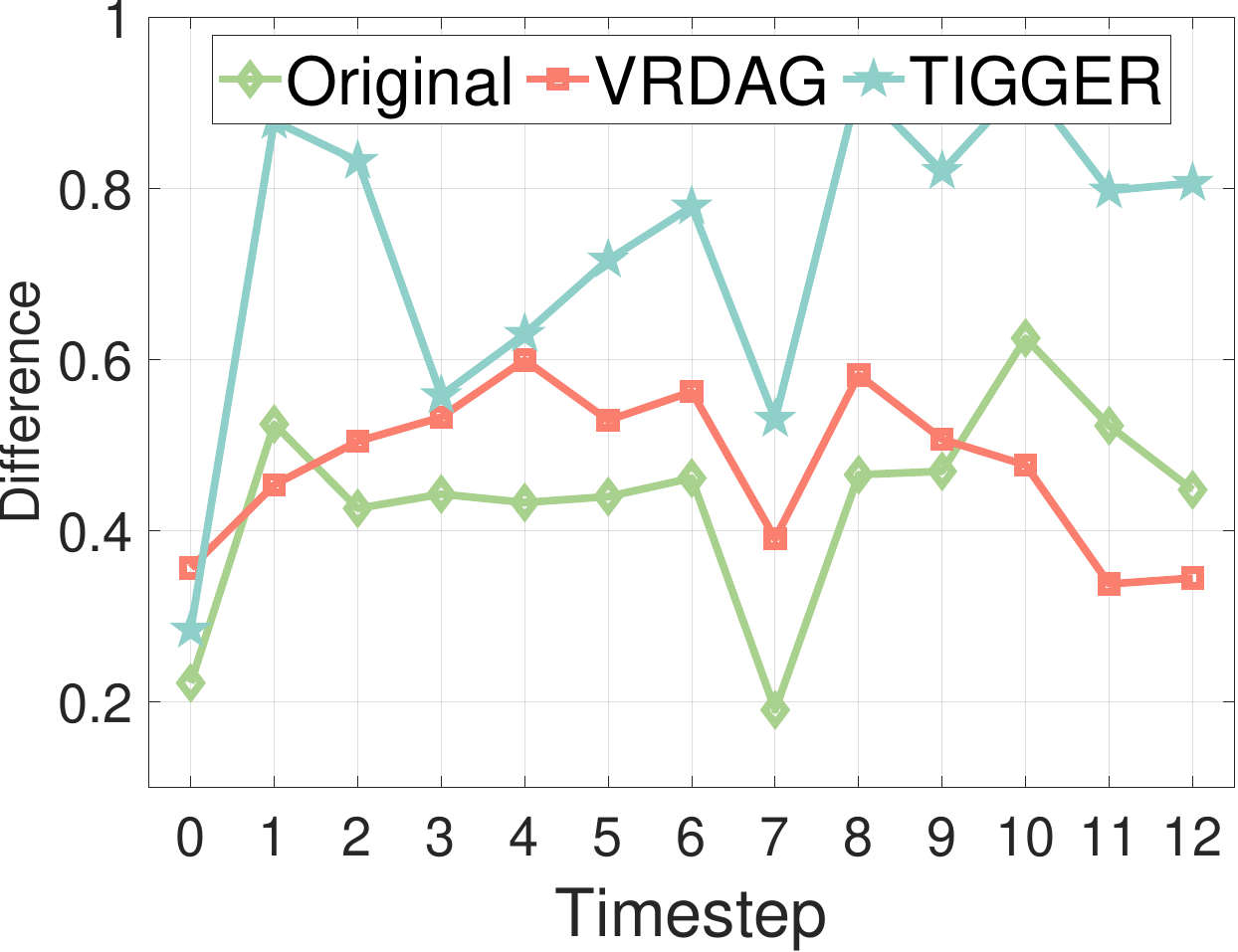}}
	\subfloat[Wiki]{
		\includegraphics[width=0.32\linewidth]{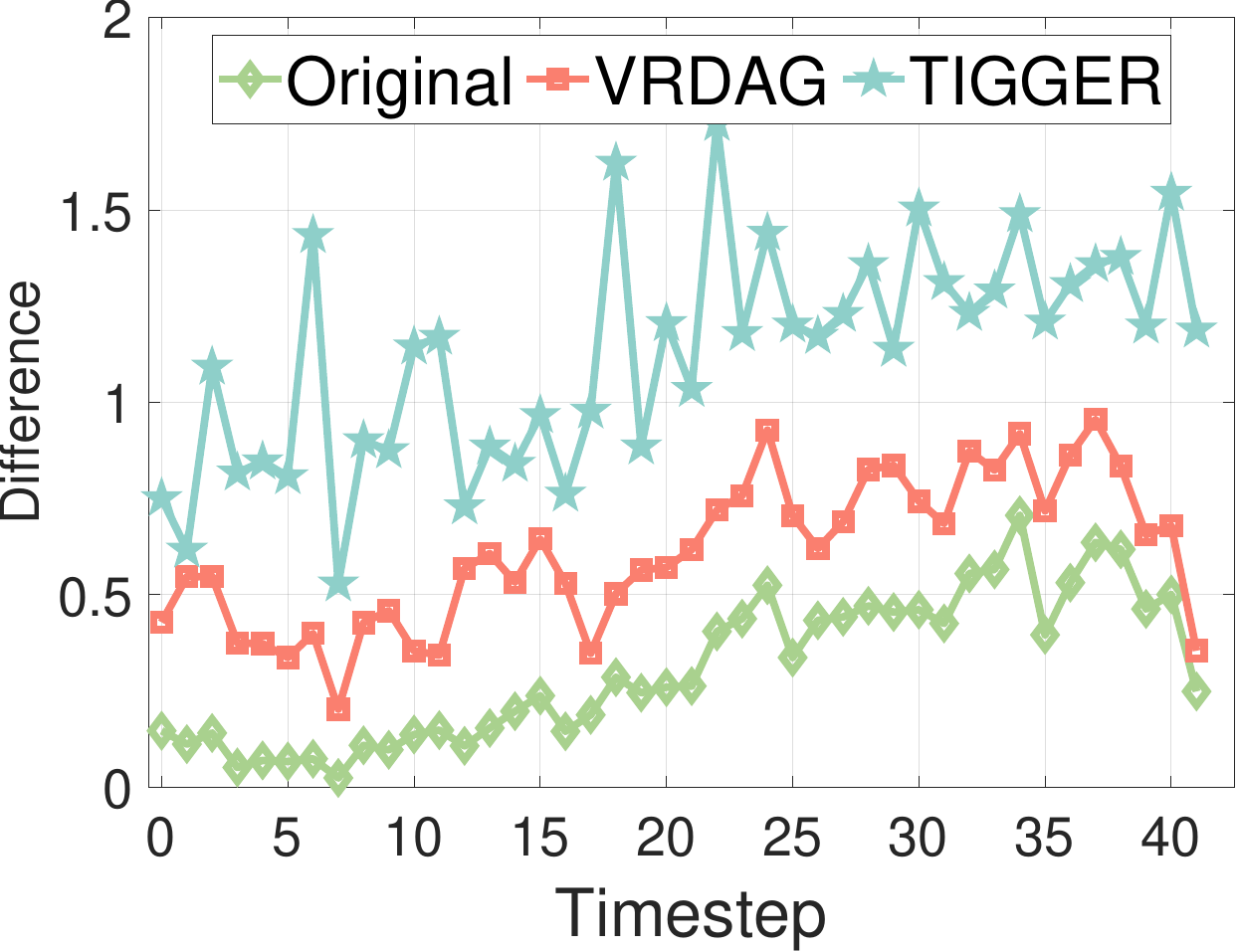}}
        \subfloat[GDELT]{
		\includegraphics[width=0.32\linewidth]{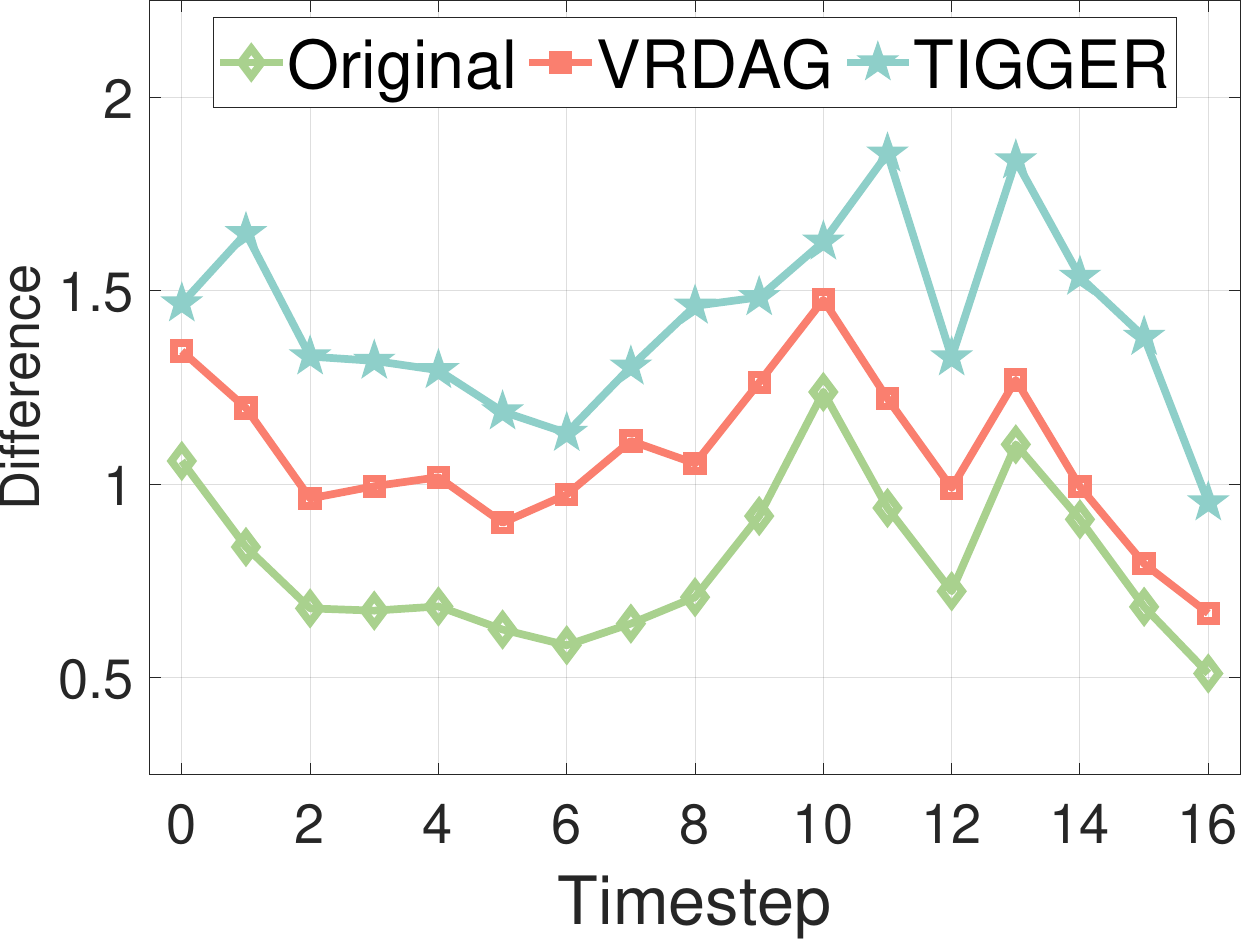}} 
        \caption{Temporal structure difference in coreness}
        \label{fig:core}
\end{figure}


\begin{figure}[t]
	\centering
	\subfloat[Email]{
		\includegraphics[width=0.32\linewidth]{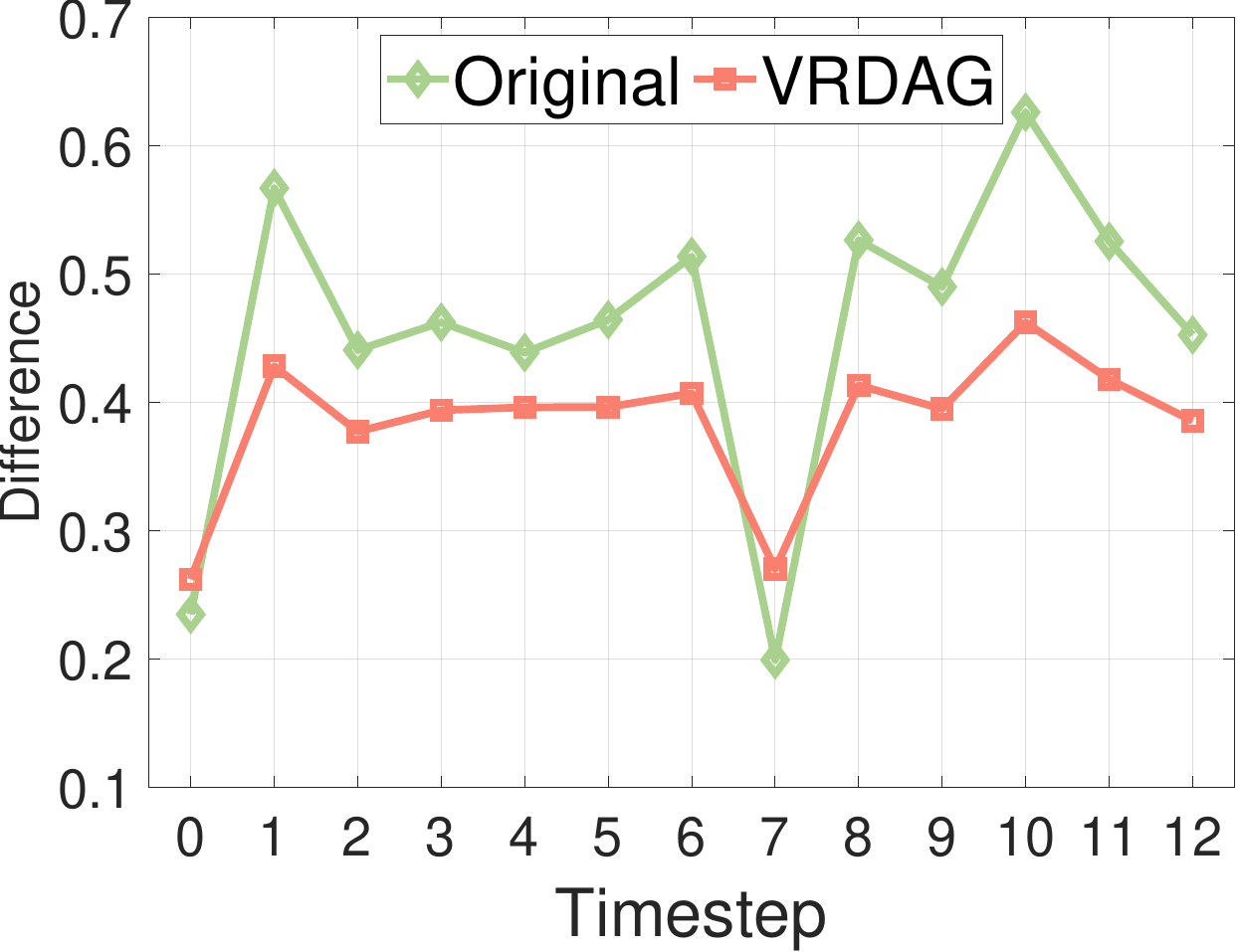}}
	\subfloat[Wiki]{
		\includegraphics[width=0.325\linewidth]{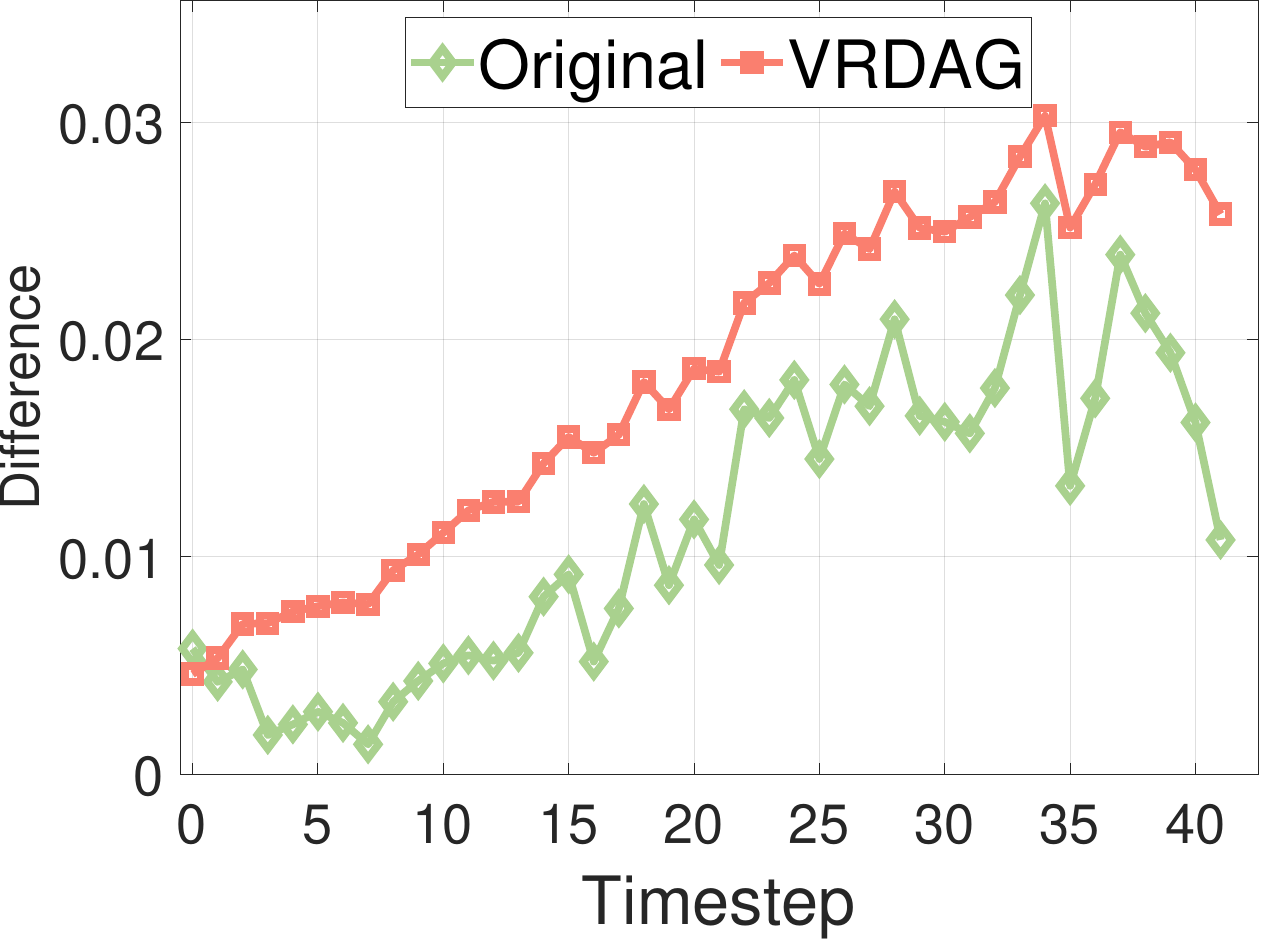}}
        \subfloat[GDELT]{
		\includegraphics[width=0.32\linewidth]{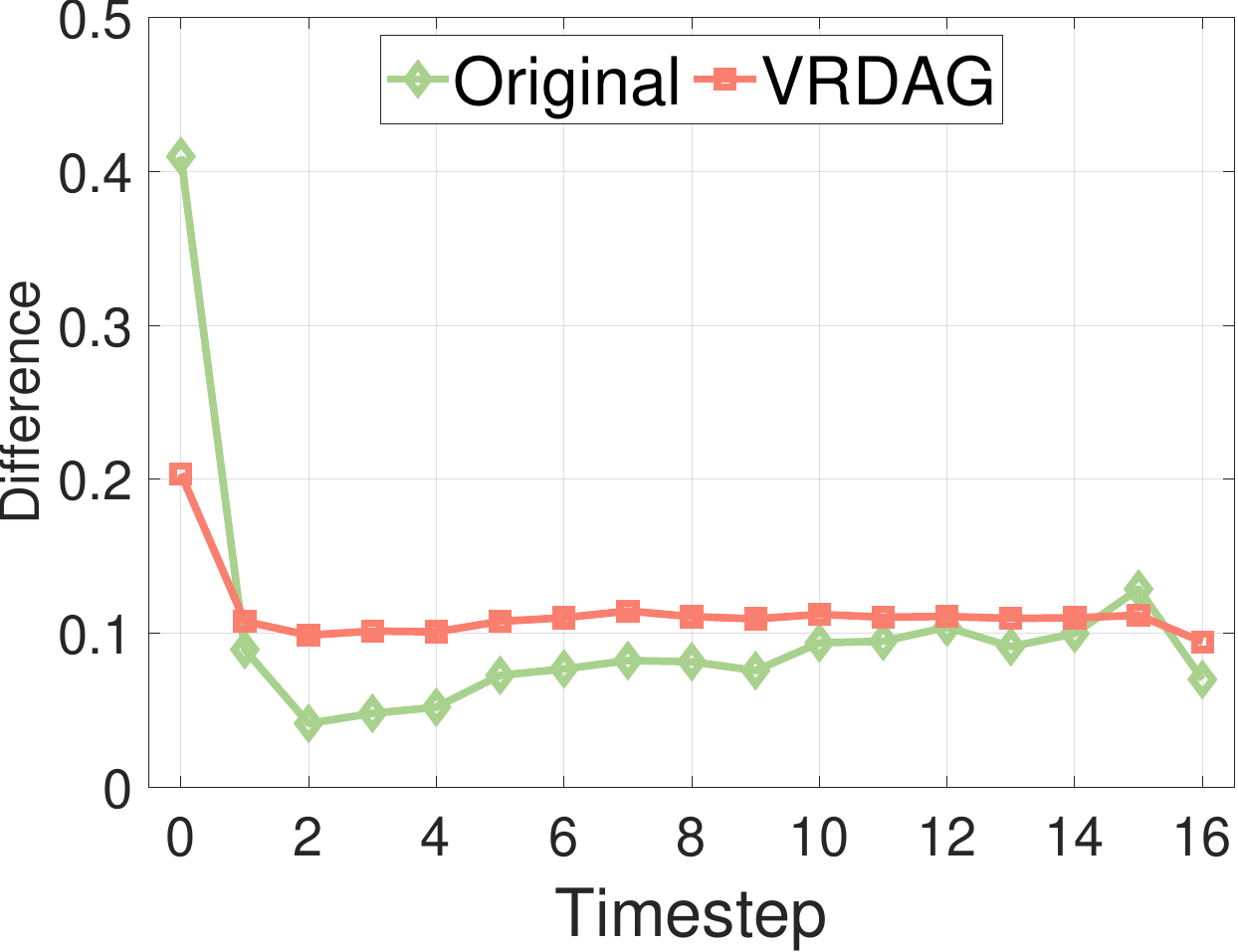}} 
        \caption{Temporal attribute difference in terms of MAE}
        \label{fig:MAE}
\end{figure}

\begin{figure}[h!]
	\centering
	\subfloat[Email]{
		\includegraphics[width=0.32\linewidth]{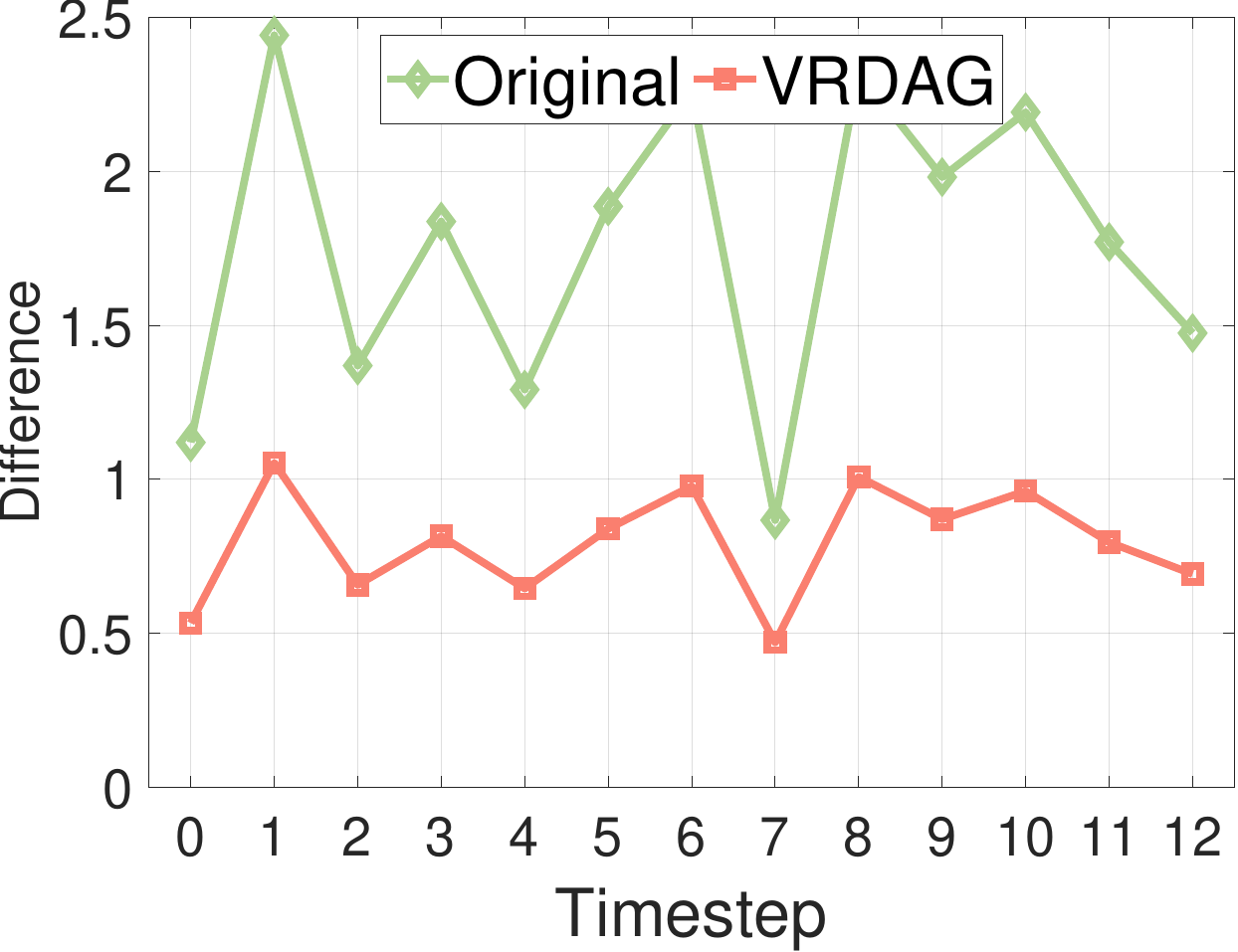}}
	\subfloat[Wiki]{
		\includegraphics[width=0.325\linewidth]{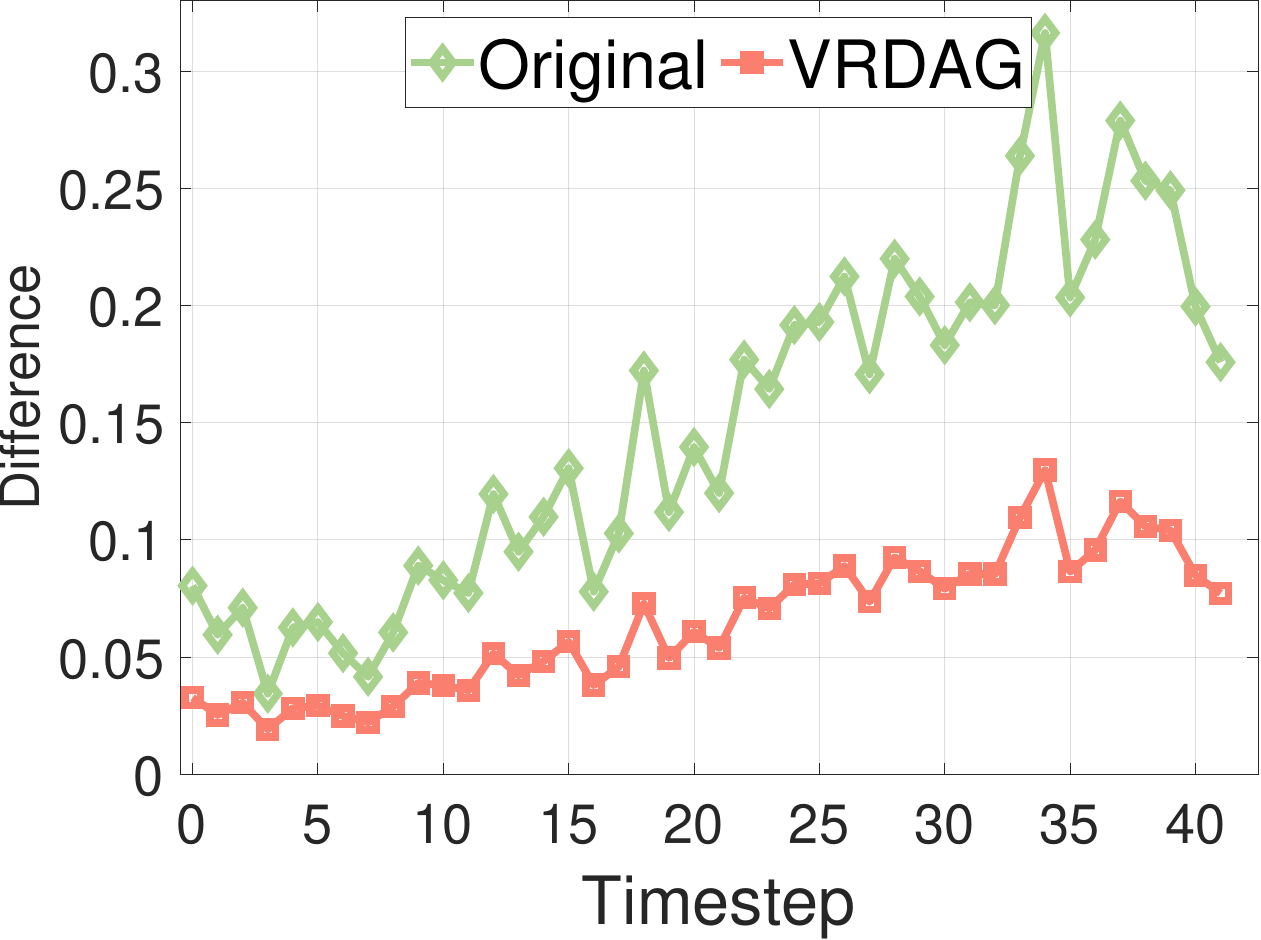}}
        \subfloat[GDELT]{
		\includegraphics[width=0.32\linewidth]{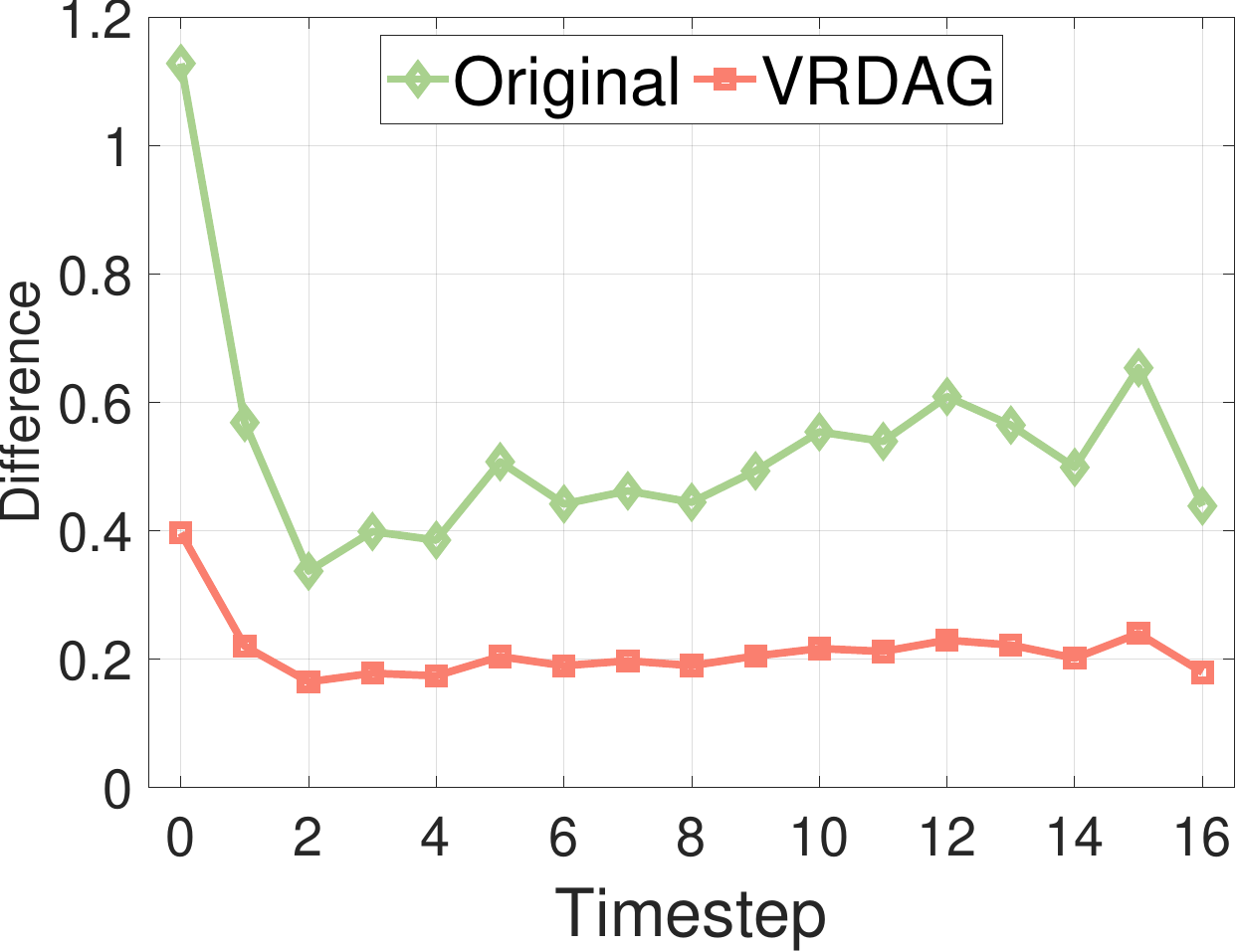}} 
        \caption{Temporal attribute difference in terms of RMSE}
        \label{fig:RMSE}
\end{figure}

\subsection{Dynamic Difference Evaluation}

To validate that VRDAG can capture the dynamic change patterns in the original graph, we compute the differences between each pair of consecutive snapshots in the graph sequence using the metrics mentioned in Section~\ref{metric}. This experiment is conducted on the Email, Wiki, and GDELT datasets, representing small, medium, and large scales, respectively. For structure difference evaluation, we compare VRDAG to the current SOTA deep generator, TIGGER. The results for the three metrics (i.e., degree, clustering coefficient, and coreness) are shown in Figures~\ref{fig:deg}-\ref{fig:core}, where VRDAG’s line plot consistently aligns more closely with the original difference sequence than TIGGER’s across various cases.
This demonstrates VRDAG’s superior ability to capture dynamic changes between consecutive snapshots of the original data. For example, considering degree differences on Wiki, the discrepancy between VRDAG and the original data remains consistently below 0.85, while TIGGER's discrepancy ranges from 1.08 to 2.87.
Regarding attribute differences, as there is no existing baseline for dynamic attributed graph generation, we directly compare VRDAG to the original graph. As shown in Figures~\ref{fig:MAE}-\ref{fig:RMSE}, VRDAG effectively models dynamic attribute changes. For instance, in terms of MAE, VRDAG captures the stable trend in attribute differences on the GDELT dataset, while it reflects an increasing trend in attribute differences over the sequence on the Wiki dataset.

\subsection{Efficiency and Scalability}

\begin{figure} [t]
	\centering
	\subfloat[Training time]{
		\includegraphics[width=0.45\linewidth]{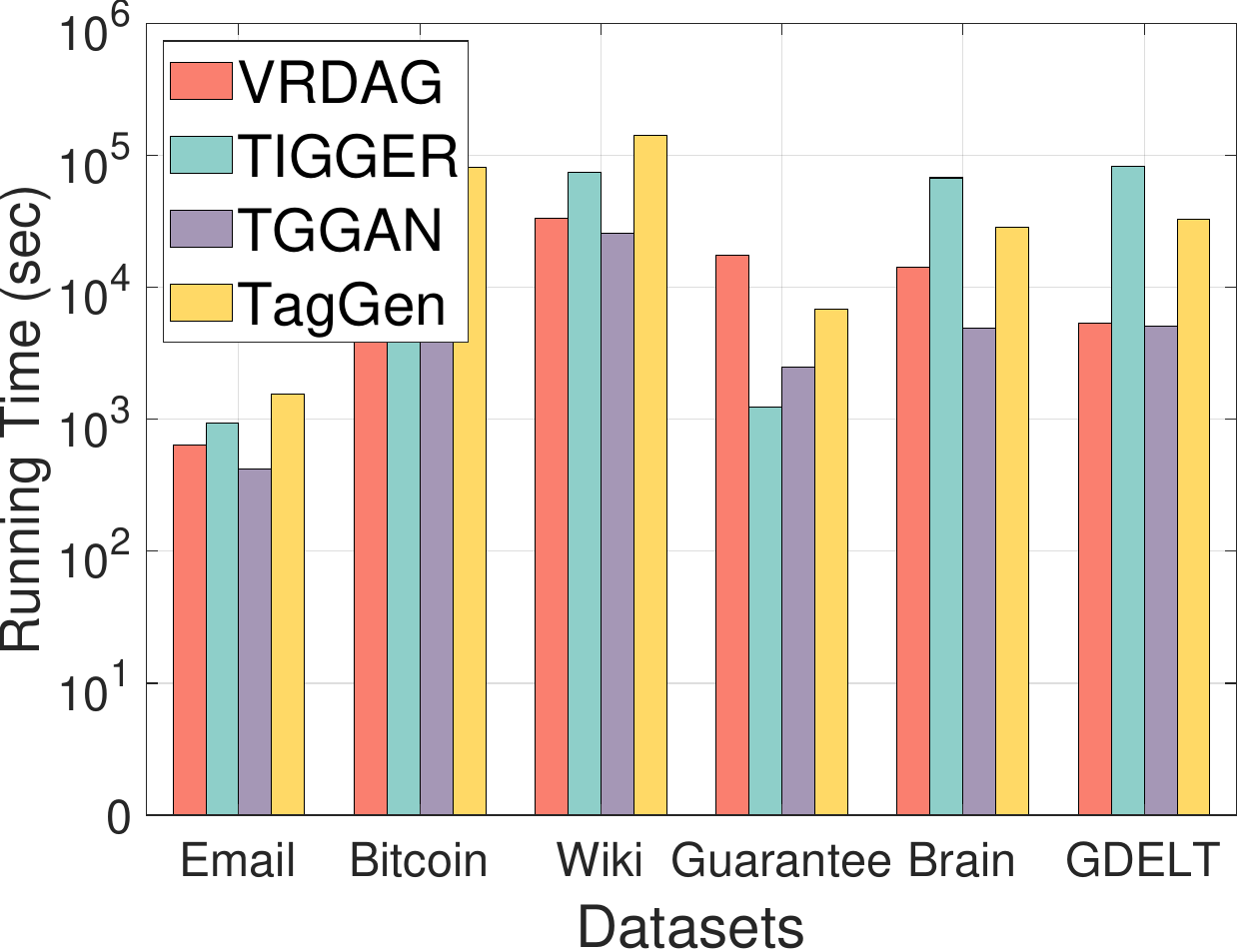}}
        \subfloat[Testing time]{
		\includegraphics[width=0.45\linewidth]{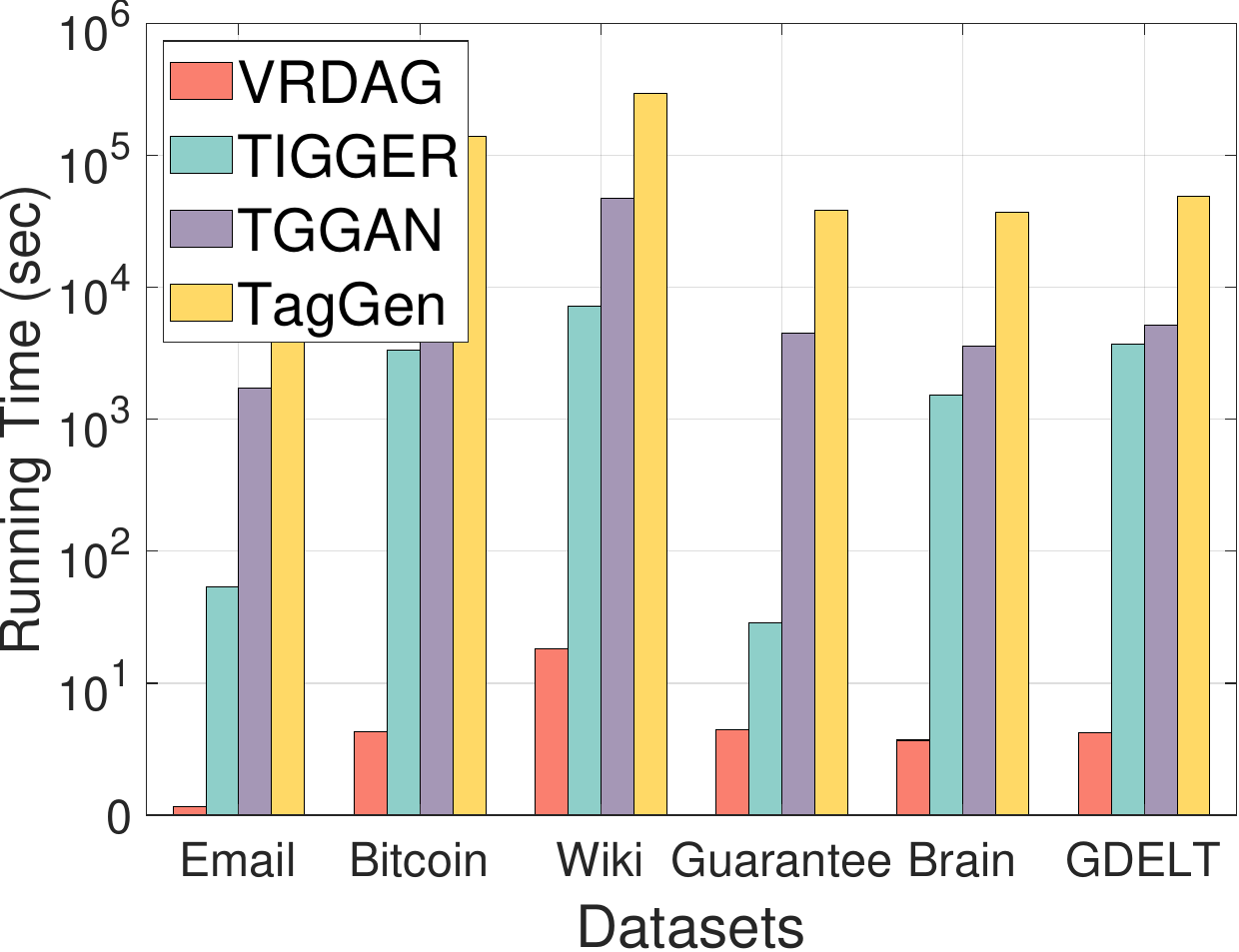} }
	\\
	\subfloat[Training time trend]{
		\includegraphics[width=0.45\linewidth]{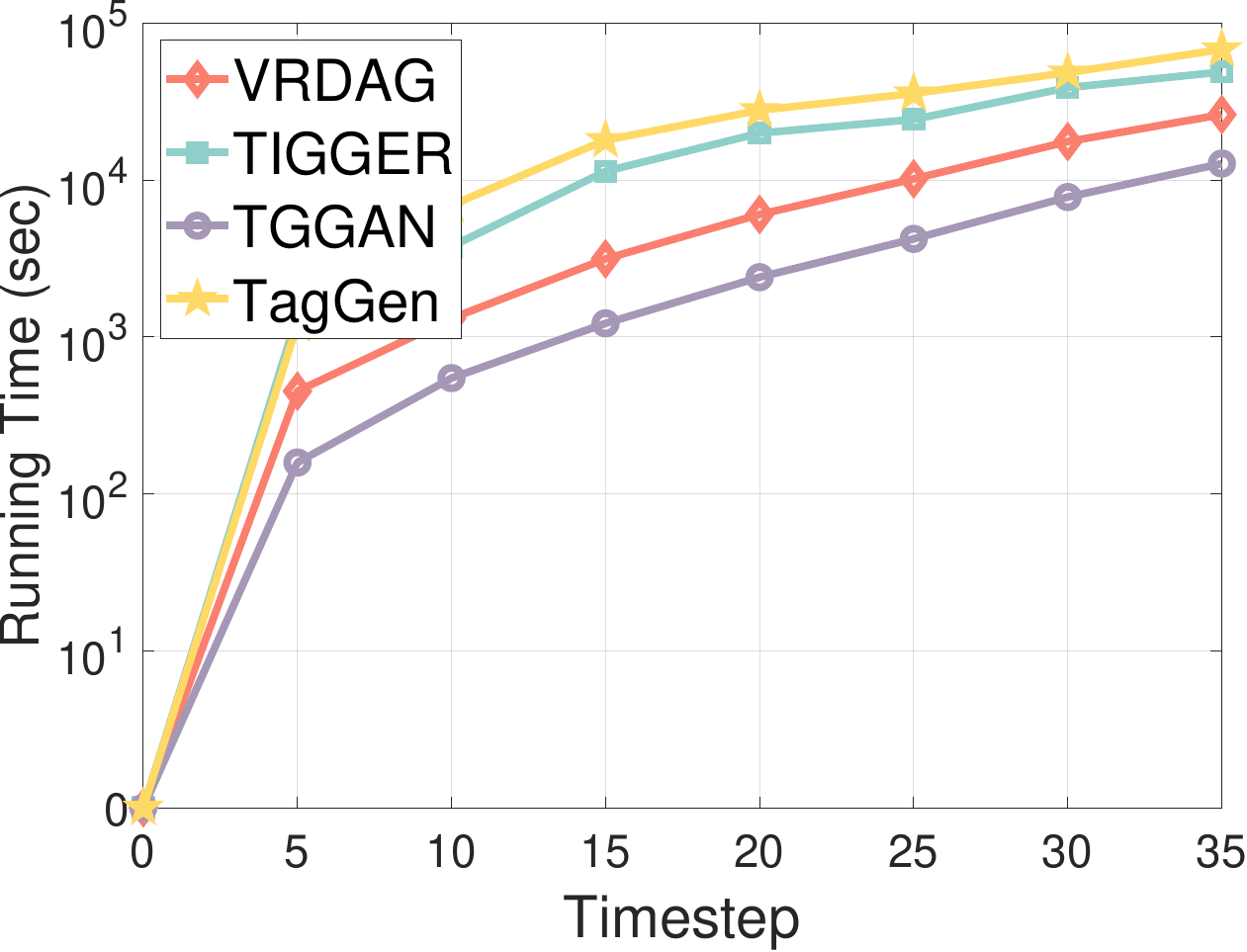}}
	\subfloat[Testing time trend]{
		\includegraphics[width=0.45\linewidth]{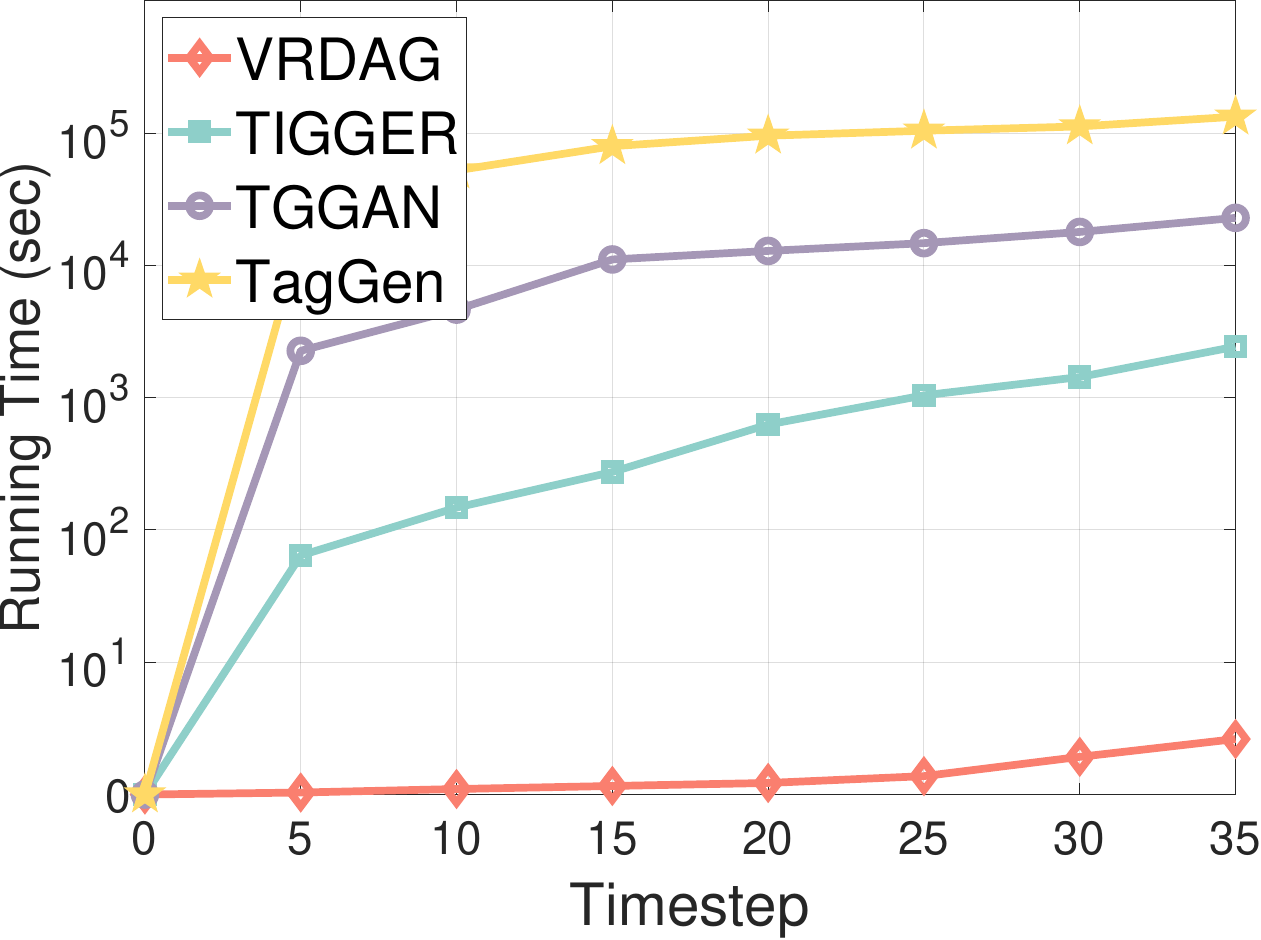}}
        \caption{Efficiency evaluation. (a) and (b) show the running time comparison results in the training and testing stages. (c) and (d) present the models' running time against timesteps in both stages on Bitcoin. The y-axis is in the log scale.}
        \label{fig:efficiency}
\end{figure}

\myparagraph{Efficiency evaluation} To demonstrate the efficiency of VRDAG in generating dynamic attribute graphs, we first report the overall running time of VRDAG, TagGen, TGGAN, and TIGGER in both the training and testing stages on four real-world datasets. Fig.~\ref{fig:efficiency}(a) shows that TGGAN needs lower training time on different datasets. However, as reported in TABLE~\ref{tab:comp}, this method cannot generate high-quality dynamic graph structures as our VRDAG. Our recurrent model achieves the second-highest training efficiency on most datasets.
In the inference phase (Fig.~\ref{fig:efficiency}(b)), VRDAG is the most efficient of all models, outperforming TIGGER by up to 2 orders of magnitude in efficiency. Additionally, our recurrent framework is up to 4 orders of magnitude faster than TagGen across different datasets.
Notably, on the Wiki dataset, our recurrent framework requires only 18.3 seconds to simulate attributed snapshots over 43 timesteps. In contrast, TIGGER takes 2 hours for generation, TGGAN requires over 13 hours, and TagGen needs more than 3 days to generate the structure. This is because TagGen utilizes a discriminator to gradually select plausible generated temporal random walks before feeding them into the merging module. As the number of sampled temporal paths increases in large dynamic networks with long timesteps, this discrimination stage significantly limits the generation speed. TGGAN further utilizes a truncated temporal random walk which can reduce the running time. TIGGER applies a pre-trained RNN to sample temporal paths, achieving relatively better efficiency compared to the above two methods. However, unlike the random walk-based methods, which require extensive path sampling and merging, our recurrent model preserves the evolving node embeddings updated at the last timestep and generates the next-step snapshot with a one-shot decoding process. This scheme considerably reduces the generation time. Additionally, we analyze the trend of time growth for each model by plotting the running time against the number of timesteps on Bitcoin dataset (Fig.~\ref{fig:efficiency}(c) and Fig.~\ref{fig:efficiency}(d)). It can be observed that while TGGAN is the most efficient during training, our framework is consistently more efficient than the three temporal random walk-based methods in the generation process when varying the lengths of snapshots. We also find that VRDAG exhibits a slower time growth rate during generation. Fig.~\ref{fig:efficiency}(d) shows that as the timestep increases from 0 to 5, the synthesis time of TIGGER rises by about 80 seconds, while the running time of TagGen surges to over 1.5$\times 10^{4}$ seconds. In contrast, our VRDAG only requires about 1 second when synthesizing the dynamic graph with 5 timesteps. When we vary the timestep from 5 to 35, the time cost of VRDAG increases by less than 3 seconds. 
This indicates the superior time scalability of our framework. 

\myparagraph{Scalability evaluation} Similar to~\cite{gupta2022tigger}, we record the growth of training and generation time against the number of temporal edges sampled from the GDELT dataset in Table~\ref{tab:scalability-train} and Table~\ref{tab:scalability-test}, respectively. During the training process, we observe that the time consumption of TagGen scales by 13.33 times as the number of edges varies from 10k to 100k. In contrast, the running time of VRDAG only scales by less than 4.9 times. During the generation process, we notice that our recurrent method outperforms the three temporal walk-based methods in efficiency by at least 1 order of magnitude when the edge number increases from 1k to 500k, demonstrating its superior scalability against graph size. When the number of sampled temporal edges scales to 500k, VRDAG generates a graph with only 3.78 seconds, outperforming TagGen, TGGAN, and TIGGER by 2 to 4 orders of magnitude in running time.

\begin{table}[t!]
\caption{Scalability against different sizes of temporal edges on GDELT dataset during training (seconds)}
\tabcolsep 8pt
\centering
\begin{tabular}{c|c|c|c|c}
\toprule
\multicolumn{1}{c|}{\#Edges} & \multicolumn{1}{c|}{1k} & \multicolumn{1}{c|}{10k} & \multicolumn{1}{c|}{100k} & \multicolumn{1}{c}{500k} \\ \midrule
TagGen &  1.33e2 & 9.15e2 & 1.22e4 &  3.04e4   \\
TGGAN &  \textbf{4.24e1} &  \textbf{3.87e2} & 2.40e3 & \textbf{4.71e3}      \\
TiGGER &  8.27e1 &  1.12e3 & 1.30e4 & 7.14e4      \\
 VRDAG &  1.47e2 & 4.77e2 & \textbf{2.33e3} & 5.07e3     \\ \bottomrule
\end{tabular}
\label{tab:scalability-train}
\end{table}

\begin{table}[t]
\caption{Scalability against different sizes of temporal edges on GDELT dataset during generation (seconds)}
\tabcolsep 8pt
\centering
\begin{tabular}{c|c|c|c|c}
\toprule
\multicolumn{1}{c|}{\#Edges} & \multicolumn{1}{c|}{1k} & \multicolumn{1}{c|}{10k} & \multicolumn{1}{c|}{100k} & \multicolumn{1}{c}{500k} \\ \midrule
TagGen &  1.38e3 & 7.06e3 & 2.46e4 &  3.93e4   \\
TGGAN &  1.95e2 &  1.08e3 & 3.65e3 & 4.92e3      \\
TiGGER &  6.43 &  56.64 & 8.30e2 & 3.49e3      \\
 VRDAG &  \textbf{3.23e-1} & \textbf{8.66e-1} & \textbf{2.29} & \textbf{3.78}     \\ \bottomrule
\end{tabular}
\label{tab:scalability-test}
\end{table}

\subsection{Case Study}

\begin{figure}[t!]
	\centering
	\subfloat[Link Prediction]{
		\includegraphics[width=0.45\linewidth]{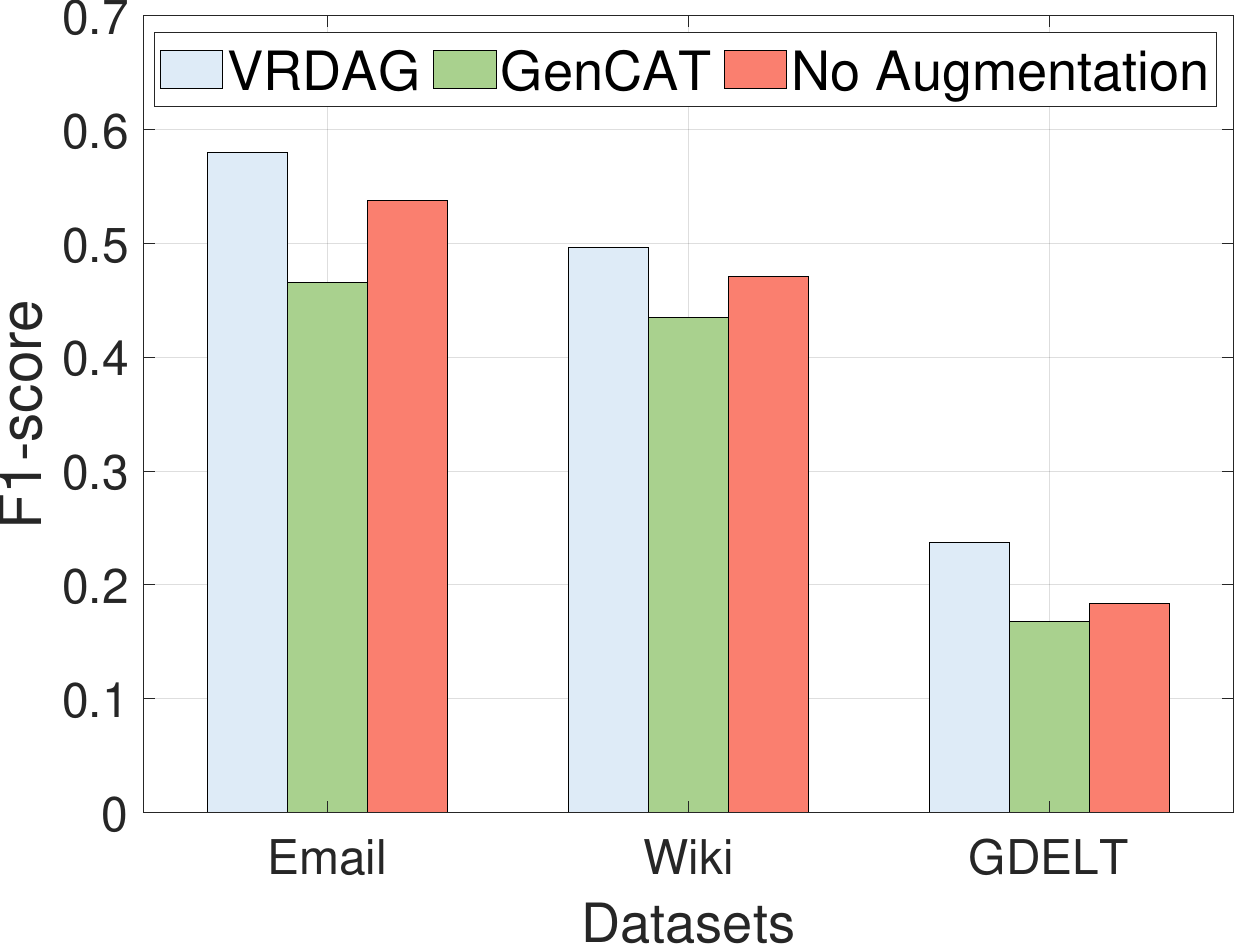}}
	\subfloat[Attribute Prediction]{
		\includegraphics[width=0.45\linewidth]{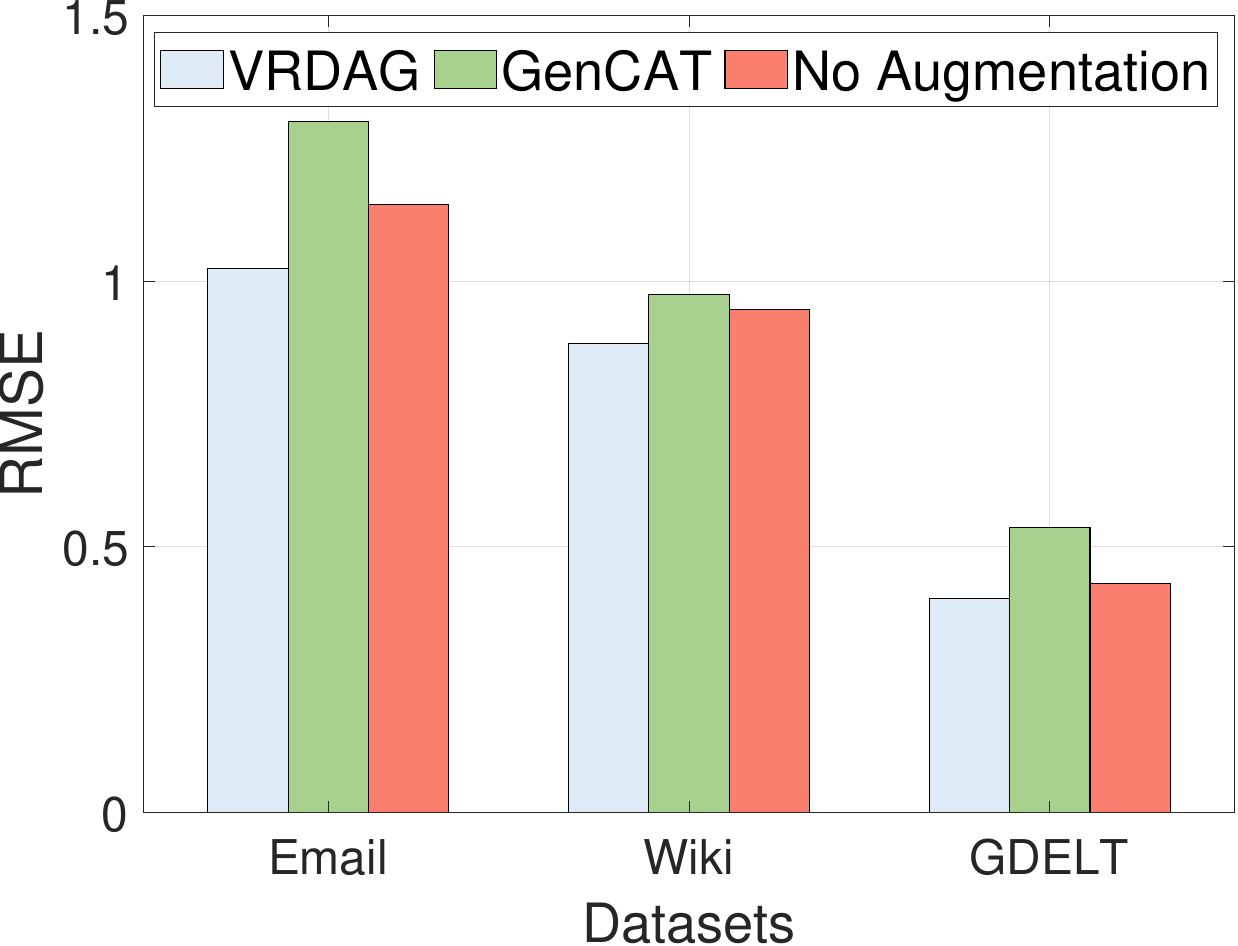}}
        \caption{Application of generated graphs on downstream tasks: Link Prediction and Attribute Prediction}
        \label{fig:down}
\end{figure}

To validate the utility of our generated graph, we apply it to data augmentation on the downstream task of forecasting the entire future graph snapshot, which can be decomposed into the link prediction task and the node attribute prediction task. The CoEvoGNN~\cite{wang2021modeling}, which is the SOTA dynamic attributed graph prediction method, is adopted as our training model. We conduct experiments on the Email, Wiki, and GDELT, which represent small, medium, and large datasets respectively. Following~\cite{wang2021modeling}, we test on predicting the final snapshot
and use the previous graph sequence as the training data. In our experiments, we train the CoEvoGNN on the original graph sequence and the synthetic graph sequence, which can be viewed as augmented data. We compare the models trained on augmented data generated by VRDAG and GenCAT. The F1-score and RMSE are employed as metrics for link prediction and attribute prediction, respectively. We repeat each experiment for 5 runs and report the average result. As shown in Figure~\ref{fig:down}, we observe that the model trained on augmented data generated by VRDAG consistently achieves superior performance than the base model without data augmentation on both tasks across three datasets. For example, on GDELT, VRDAG improves the F1-score by 6.3\% compared to the base model in link prediction, while for attribute prediction, VRDAG achieves a 0.12 lower RMSE. Additionally, we find that augmented data simulated by GenCAT would deteriorate model performance on both tasks. This is because the snapshots generated by the static GenCAT are independent, which fails to capture the original node behaviors across the sequence, thus harming the training process.

\section{Related work}
\label{sec:rel}

In this section, we summarize the related works regarding relational data generation, static/dynamic graph generation, and the variational recurrent model (See Appendix B-A~\cite{onlineapp}).

\myparagraph{Relational data generation} In recent years, generating synthetic relational data has attracted significant interest in database research. To benchmark data management solutions, Arasu et al.~\cite{arasu2011data} study the problem of synthetic databases and propose a declarative approach for specifying data characteristics using cardinality constraints. This work is later extended in Hydra~\cite{sanghi2018hydra} to incorporate dynamism in the generation process. After that, PiGen~\cite{sanghi2022projection} expands the scope of the supported constraints to the general projection operator, leading to more satisfactory database generation. For practical cloud database evaluation, Yang et al.~\cite{yang2022sam} consider data generation from query workloads with supervised autoregressive models. Motivated by scalability testing for database systems, the Dataset Scaling Problem (DSP)~\cite{tay2013upsizer} which aims to generate scaled relational tables with the given factor has been introduced. Dscaler~\cite{zhang2016dscaler} further considers non-uniform scaling (nuDSP), allowing tables to scale by different factors.


\myparagraph{Static graph generation}
The generation of realistic graph data is a long-standing research problem. Traditional graph generators have focused on various families of random graph models, such as the Barabási-Albert model~\cite{albert2002statistical}, 
stochastic block models~\cite{airoldi2008mixed}, Kronecker graph models~\cite{leskovec2010kronecker}, and exponential random graph models~\cite{robins2007introduction}. To generate large-scale graphs for graph processing system benchmarking, TrillionG~\cite{park2017trilliong} proposes a scope-based model by generalizing the random graph model~\cite{albert2002statistical}. FastSGG~\cite{wang2021fastsgg} can efficiently generate high-quality social networks using user-defined configurations. However, the graph generators mentioned above are limited to reproducing specific statistical properties and fall short of capturing complex graph data distributions. To address this, data-driven deep learning methods have been introduced. VAE-based methods~\cite{kipf2016variational,ma2018constrained} parameterize variational autoencoder with graph neural network (GNN)~\cite{liang2023ba} to generate individual entries in adjacency matrix independently. RNN-based generation models, such as GraphRNN~\cite{you2018graphrnn} and GRAN~\cite{liao2019efficient}, model the probability of generating edges at each step using a specific output distribution conditioned on the already generated graph. Bojchevski et al.~\cite{bojchevski2018netgan} employ Generative Adversarial Networks (GANs) to learn the distribution of biased random walks over the input graph for network generation, achieving superior performance. 
Most existing static graph generators primarily focus on the generation of structures while ignoring node attributes. Pfeiffer et al.~\cite{pfeiffer2014attributed} introduce the Attributed Graph Model (AGM) framework to jointly model network structure and vertex attributes. Nonetheless, the AGM is limited in computing edge probabilities conditioned on correlated attributes and lacks the ability to synthesize accurate node attributes. To address attribute generation, Largeron et al.~\cite{largeron2015generating} propose ANC, which generates numeric node attribute values following a normal distribution. While simple, this method is not suitable for generating real-world graphs. Recently, Maekawa et al.~\cite{maekawa2023gencat} introduce GenCAT, which can generate attributes following user-specified distributions, providing greater flexibility. 

\myparagraph{Dynamic graph generation}
Generating temporal graph data is a crucial problem, considering that many real-world networks exhibit inherent dynamics. Traditional methods for generating dynamic graphs mainly focus on modeling the evolution of nodes and edges over time~\cite{holme2013epidemiologically,rocha2013bursts,vestergaard2014memory}. For instance, Holme~\cite{holme2013epidemiologically} first generates a static graph from a random graph model and then assigns active interval durations to each link from a truncated power-law distribution. 
Vestergard et al.~\cite{vestergaard2014memory} utilize temporal memory effects to model the active state of nodes and edges. 
To further incorporate the high-order structure evolution patterns, Zeno et al.~\cite{zeno2021dymond} propose a motif-based approach, assuming that each type of motif follows a time-independent exponentially distributed arrival rate. 
However, these methods are often hand-engineered and require prior knowledge or assumptions, lacking the ability to directly learn from the data. To address this, Zhou et al. propose TagGen~\cite{zhou2020data}, a deep generative framework based on temporal random walks that jointly extract structural and temporal context information from dynamic networks. To capture time-validity constraints and joint distribution of time and topology, TGGAN~\cite{zhang2021tg} is developed as a successor of TagGen.
Gupta et al.~\cite{gupta2022tigger} further propose TIGGER to make the walk-based methods more scalable. 


\section{Conclusion}
\label{sec:con}

In this paper, we propose VRDAG, a data-driven variational recurrent framework, for efficient dynamic attributed graph generation. In particular, we first introduce a bi-flow graph encoder to preserve relational knowledge in the newly generated snapshot. Then, a recurrence updater is applied to update hidden node states. Next, a learned prior distribution is used to generate temporal latent variables based on the hidden states of early graphs. This technique can capture dependencies between and within topological and node attribute evolution processes. Sampled latent variables are subsequently fed into a factorized decoder with a MixBernoulli sampler and an attentive attribute decoder to generate the new snapshot. We obtain the final graph sequence by recurrently synthesizing snapshots in this manner. Extensive experimental results show VRDAG's effectiveness in generating high-quality dynamic attributed networks. Moreover, the efficiency evaluation proves that our recurrent model significantly reduces generation time compared to existing state-of-the-art deep generative methods.

\section*{Acknowledgment}
Ying Zhang was supported by Alibaba K24-0448-011. Xiaoyang Wang was supported by ARC DP230101445 and DP240101322. Dawei Cheng was supported by the National Science Foundation of China (62472317)

\bibliography{refs}

\end{document}